\documentclass[prd,superscriptaddress,eqsecnum,nofootinbib,notitlepage,twocolumn]{revtex4-1}

\usepackage{comment}
\usepackage{enumerate}
\usepackage{mathrsfs}
\usepackage{units}
\usepackage{hyperref,graphicx,amsfonts,amssymb,amsthm,amsmath,psfrag}
\usepackage[toc,page]{appendix}
\usepackage{subfigure}
\usepackage{slashed}
\usepackage{tikz}
\usepackage{booktabs}
\usepackage{siunitx}

\usepackage[normalem]{ulem}	
\newcommand\redout{\bgroup\markoverwith
{\textcolor{red}{\rule[.5ex]{8pt}{0.8pt}}}\ULon}
\usepackage{xcolor}
\usepackage{color}
\usepackage{todonotes}

\usetikzlibrary{decorations.markings,snakes}
\tikzset{
  fermionline/.style={line width=1pt,postaction={decorate},
    decoration={markings,
      mark=at position 0.5 with {\draw[-stealth] (0,0)--(2pt,0);}}},
  bosonline/.style={line width=1pt,decorate,
    decoration={snake,amplitude=1,segment length=4}},
  higgsline/.style={line width=1pt,dashed}
}

\newcommand{\be}{\begin{equation} }
\newcommand{\ee}{\end{equation}}
\newcommand{\cL}{\mathcal{L}}
\newcommand{\e}{\mathrm{e}}
\newcommand{\Tr}{\mathop\text{Tr}}
\newcommand{\MSb}{$\overline{\text{MS}}$}

\begin{document}
\title{Living beyond the edge: Higgs inflation and vacuum metastability}
\author{Fedor Bezrukov}
\email{fedor.bezrukov@uconn.edu}
\affiliation{Physics Department, University of Connecticut, Storrs, Connecticut 06269-3046, USA}
\affiliation{RIKEN-BNL Research Center, Brookhaven 
National Laboratory, Upton, New York 11973, USA}
\affiliation{CERN, CH-1211 Geneve 23, Switzerland}
\author{Javier Rubio}
\email{javier.rubio@epfl.ch}
\affiliation{Institut de Th\'{e}orie des Ph\'{e}nom\`{e}nes Physiques, 
\'{E}cole Polytechnique F\'{e}d\'{e}rale de Lausanne, CH-1015 Lausanne, 
Switzerland}
\author{Mikhail Shaposhnikov}
\email{mikhail.shaposhnikov@epfl.ch}
\affiliation{Institut de Th\'{e}orie des Ph\'{e}nom\`{e}nes Physiques, 
\'{E}cole Polytechnique F\'{e}d\'{e}rale de Lausanne, CH-1015 Lausanne, 
Switzerland}
\date{\today}

\begin{abstract}
The  measurements of the Higgs mass and top Yukawa coupling indicate that we live in a very special universe, at the edge of the absolute stability of the electroweak vacuum.  If fully stable, the Standard Model (SM) can be extended all the way up to the inflationary scale and the Higgs field, nonminimally coupled to gravity with strength $\xi$, can be responsible for inflation.  We  show that the successful Higgs inflation scenario can also take place if the SM vacuum is not absolutely stable. This conclusion is based on two effects that were overlooked previously. The first one is associated with the effective renormalization of the SM couplings at the energy scale $M_P/\xi$, where $M_P$ is the Planck scale. The second one is a symmetry restoration after inflation due to high temperature effects that leads to the (temporary) disappearance of the vacuum at Planck values of the Higgs field.

\end{abstract}
\maketitle

\section{Introduction}

One of the most interesting questions in particle physics and cosmology is the relation between the properties of elementary particles and the structure of the Universe. Some links are provided by dark matter and the baryon asymmetry of the Universe. A number of constraints on hypothetical new particles can be also derived from big bang nucleosynthesis.

\begin{figure}
\centering
\includegraphics[scale=1]{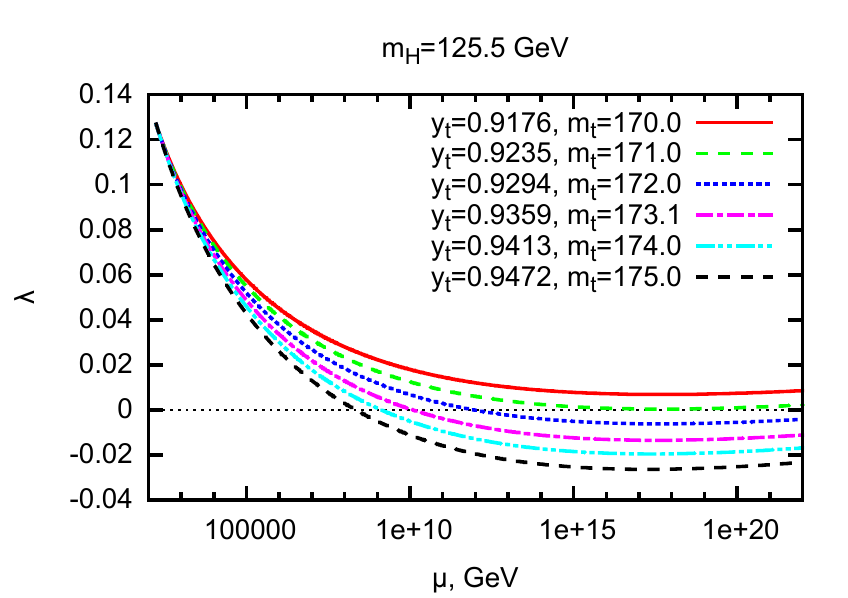}
\caption{Renormalization group running of the Higgs self-coupling for several
values of the top quark Yukawa coupling (top pole mass) and fixed $125.5$ GeV Higgs boson mass.}
\label{fig:0}
\end{figure}

\begin{figure}
\centering
\includegraphics[scale=1]{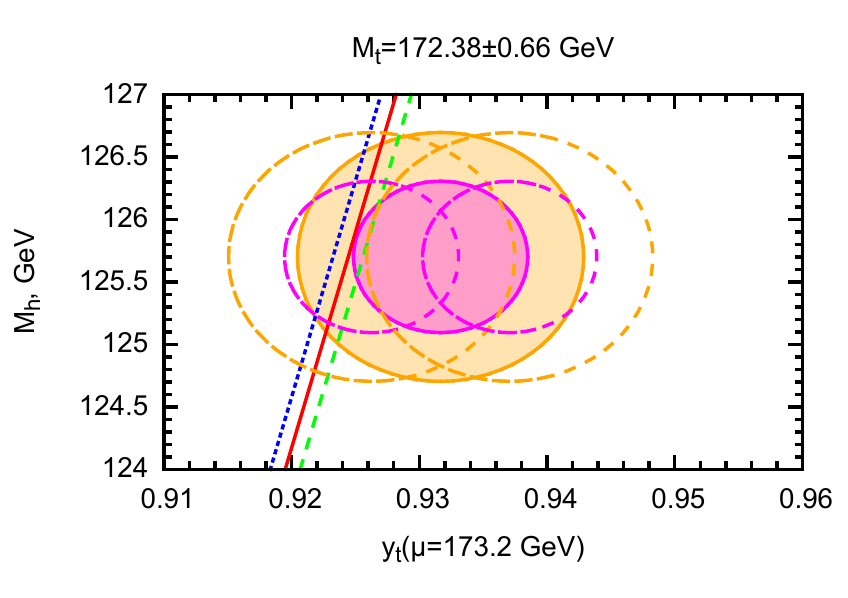}
\caption{(color online) 
The figure, taken from Ref.~\cite{Bezrukov:2014ina}, shows the borderline between the regions of absolute stability and metastability of the SM vacuum on the plane of the Higgs boson mass and top quark Yukawa coupling in the \MSb~~ scheme taken at $\mu=173.2$ GeV. The diagonal line stands for the critical value of the top Yukawa coupling $y_t^{\rm crit}$ as a function of the Higgs mass and the dashed lines account for the uncertainty associated to the error in the strong coupling constant $\alpha_s$. The SM vacuum is absolutely stable to the left of these lines and metastable to the right. The filled ellipses correspond to experimental values of $y_t$ extracted from the latest CMS determination \cite{CMS} of the Monte Carlo top quark mass $M_t = 172.38 \pm 0.10\, {\rm (stat)} \pm 0.65\, {\rm(syst)}\, {\rm GeV} $, if this is identified with the pole mass. 
Dashed ellipses encode the shifts associated to the ambiguous relation between pole and Monte Carlo masses. The ellipses are displaced to the right if other determinations of the Monte-Carlo top mass are used, 
$M_t = 173.34 \pm 0.27\, {\rm (stat)} \pm 0.71\, {\rm(syst)}\, {\rm GeV} $
and
$M_t = 174.34 \pm 0.37\, {\rm (stat)} \pm 0.52\, {\rm (syst)}\, {\rm GeV}$ 
coming respectively from  the combined analysis of ATLAS, CMS, CDF, and D0 data (at 8.7~fb$^{-1}$ of Tevatron Run II) \cite{ATLAS:2014wva} and from the CDF and D0 combined analysis of Run I and Run II of Tevatron \cite{Tevatron:2014cka}.}\label{fig:1}
\end{figure}

\begin{figure}
\centering
\includegraphics[scale=1]{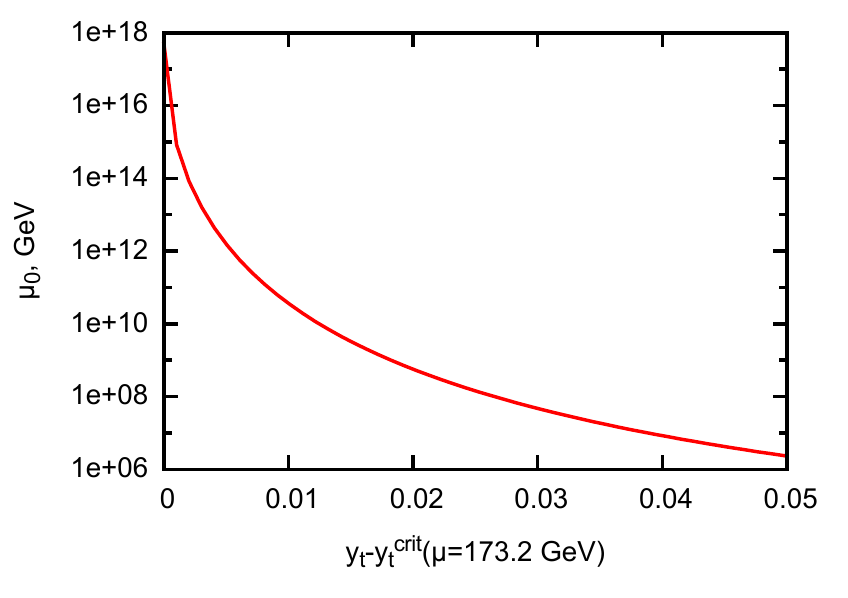}
\caption{Energy scale $\mu_0$ where the Higgs self-coupling becomes negative as a function of the deviation of the top Yukawa coupling $y_t$ from the critical value $y_t^{\rm crit}$. Adapted from Ref.~\cite{Bezrukov:2014ina}.}\label{fig:2}
\end{figure}

The properties of the recently discovered Higgs boson \cite{Aad:2012tfa,Chatrchyan:2012ufa} suggest an additional and intriguing connection. Among the many different values that the Higgs mass could have taken, nature has chosen one that  allows us to extend the Standard Model (SM) all the way up to the Planck scale while staying in the perturbative regime. The behavior of the Higgs self-coupling $\lambda$ is quite peculiar: it decreases with energy to eventually arrive to a minimum at Planck scale values and starts increasing thereafter, see~Fig.~\ref{fig:0}.  Within the experimental and theoretical uncertainties\footnote{The largest uncertainty comes from the determination of the top Yukawa coupling. Smaller uncertainties are associated with the determination of Higgs boson mass and the QCD gauge coupling $\alpha_s$. See Refs.~\cite{Bezrukov:2012sa,Degrassi:2012ry,Buttazzo:2013uya} for the most refined treatments and Ref.~\cite{Bezrukov:2014ina}  for a review.}, the Higgs coupling may stay positive all way up to the Planck scale, but it may also cross zero at some scale $\mu_0$,  which can be as low as $10^8$ GeV, see~Figs.~\ref{fig:1} and \ref{fig:2}. If that happens, our Universe becomes unstable\footnote{The determination of the lifetime of the Universe  is a rather subtle issue that strongly depends on the high energy completion of the SM. As shown in Refs.~\cite{Branchina:2013jra,Branchina:2014usa,Branchina:2014rva}, if the gravitational corrections are such that the resulting effective potential lies above/below the SM one, the lifetime of our vacuum will be notably larger/smaller than the age of the Universe.}. 

The  $0-3\sigma$ compatibility of the data with vacuum instability is one of the recurrent arguments for invoking new physics beyond the Standard Model. In particular, it is usually stated that the minimalistic Higgs inflation scenario \cite{Bezrukov:2007ep}, in which the Higgs field is nonminimally coupled to gravity with strength $\xi$,  cannot take place if the Higgs self-coupling becomes negative at an energy scale below the inflationary scale. 

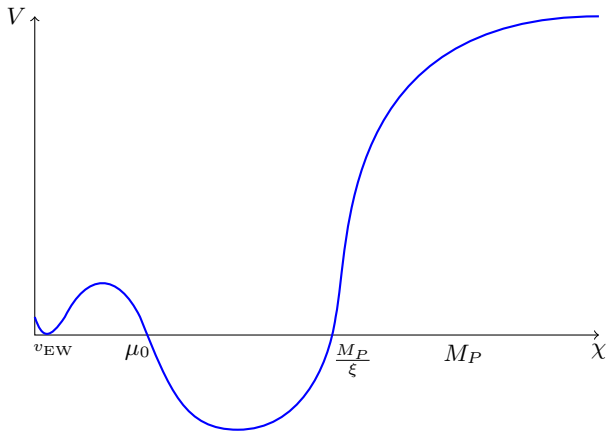
\begin{figure}
\centering
\begin{tikzpicture}
\draw[<->] (0,4) node[left] {$V$} -- (0,-0.24) -- (7.5,-0.24) node[below] {$\chi$};
\draw[thick,blue]
(0,0) .. controls (0.1,-0.3) and (0.2,-0.3) ..
(0.4,0) .. controls (0.7,0.6) and (1.1,0.6) ..
(1.4,0) .. controls (1.8,-1) and (2,-1.5) ..
(2.7,-1.5) .. controls (3.5,-1.5) and (3.9,-0.7) ..
(4,0) .. controls (4.2,1) and (4,4) ..
(7.5,4);
\node[below] at (0.25,-0.25) {$\scriptstyle v_{\text{EW}}$};
\node[below,xshift=-0.4em] at (1.5,-0.25) {$\mu_0$};
\node[below,xshift=0.7em] at (4,-0.25) {$\frac{M_P}{\xi}$};
\node[below] at (5.7,-0.25) {$M_P$};
\end{tikzpicture}
\caption{Sketch of the effective Higgs inflation potential for the scenario considered in this paper. It contains an inflationary plateau at $\chi \gtrsim M_P$ and two minima. The shallowest and narrowest one is the standard electroweak vacuum $v_{EW}$. The deepest and widest one is generated by the interplay between the instability of the Higgs self-coupling beyond the scale $\mu_0$ and the renormalization effects appearing at the scale $M_P/\xi$. }\label{cartoon}
\end{figure}

We will show in this paper that Higgs inflation is possible \textit{even if the SM vacuum is not absolutely stable}. Specifically, we will demonstrate that the renormalization effects at the scale $M_P/\xi$ can bring the Higgs self-coupling $\lambda$ to positive values in the inflationary domain. If that happens, inflation will take place with the usual chaotic initial conditions and the fate of the Universe will be inevitably determined by the subsequent evolution.  
At the end of the exponential expansion, the Higgs field will start to oscillate around the bottom of the potential, which, contrary to the tree-level case, displays two minima, with the wider and deeper one located at very large Higgs values (see~Fig.~\ref{cartoon}). The  energy stored in the Higgs field right after the end of inflation {\em highly exceeds} the height of the barrier separating the minima.  We will see that this  leads to an interesting phenomenon. The oscillations of the Higgs field  induce nonperturbative particle production, and eventually, reheat the Universe.  The shape of the potential changes due to  finite-temperature/medium effects.  If the reheating temperature is sufficiently high, the symmetry gets restored and the extra minimum at large values of the Higgs field (temporally) disappears.  The Higgs field rolls down the new potential and settles down in the electroweak vacuum.  With the evolution of the Universe, the temperature decreases and the minimum at large field values reappears. However, since the probability of tunneling to the energetically more favorable state
is completely negligible, the scalar field gets trapped near the true SM minimum and stays there until the present time.

The paper is organized as follows. In Sec. \ref{sec:gen} we review the Higgs inflation model and the self-consistent approach to quantum corrections and higher-dimensional operators.  The general arguments of Sec. \ref{sec:gen} are quantified in Sec. \ref{sec:hig}, where we discuss the contribution of the finite parts of counterterms to the effective potential and formulate the renormalization group equations for the coupling constants associated to them.  In Sec. \ref{sec:infl}  we explain how the renormalization effects can allow for inflation to happen {\em even if our vacuum is metastable}. The temperature corrections to the effective potential are computed in Sec. \ref{sec:eff}, where we determine the temperature $T_+$ at which the extra minimum at large field values disappears. Section \ref{sec:preh0}  is devoted to the study of reheating in noncritical and critical Higgs inflation and the estimation of the reheating temperature to be compared with $T_+$. Finally, the conclusions are presented in Sec. \ref{sec:con}. 

\section{The general framework}\label{sec:gen}

Higgs inflation \cite{Bezrukov:2007ep} is based on the observation that the Higgs field nonminimally coupled to gravity can give rise to inflation\footnote{The modifications of Einstein's theory of Relativity by non-minimal couplings have been widely studied in the literature (see for instance Refs.~\cite{Ad82,Mi77,Smo79,Ze79,SalopekBardeenBond}).}. The relevant part of the Lagrangian is given by
\be
\label{lagr}
{\cal L}= \left(\frac{M_P^2}{2} + \xi H^\dagger H\right)  R + g^{\mu\nu}(D_\mu H)^\dagger (D_\nu H) - U(H^\dagger H)\,,
\ee
with $R$ the Ricci scalar and
\begin{equation}\label{potJ}
U(H^\dagger H)=\lambda \left(H^\dagger H- \frac{v_{EW}^2}{2}\right)^2\,
\end{equation}
the usual Higgs potential.  The nonminimal coupling $\xi$ is assumed to take some intermediate value in the range $1\ll \xi\ll M_P^2/v^2_{EW}$ with $v_{EW}$ the vacuum expectation value of the Higgs field and $M_P=2.435 \times 10^{18}$~GeV the reduced Planck mass. The value of $\xi$ will be specified more precisely below. For sufficiently large values of the Higgs field (namely $h\gg M_P /\sqrt{\xi}$), the dimensionful parameters $v^2_{EW}$ and $M_P^2$ in the Lagrangian can be neglected and the theory becomes approximately scale invariant. This asymptotic symmetry will play a central role in the further developments.

The analysis of inflation is more easily performed in the Einstein frame \cite{GarciaBellido:2008ab,Bezrukov:2008ut}. When written in terms of a canonically normalized field $\chi$, the conformally transformed Higgs potential becomes asymptotically flat\footnote{The asymptotic shift symmetry at large field values is the Einstein-frame manifestation of the approximate scale invariance we started with.}
\begin{equation}\label{potentialE1}
 V(\chi)\simeq
\begin{cases} 
      \frac{\lambda}{4}(\chi^2-v_{EW}^2)^2\,, & \chi\ll \frac{M_P}{\xi}\,, \\
 \frac{\lambda M_P^4}{4\xi^2}\left[1-e^{-\frac{\sqrt{2/3}\chi}{M_P}}\left(1+\frac{\xi v_{EW}^2}{M_P^2}\right)\right]^2\,, & \chi\gg \frac{M_P}{\xi}\,.
   \end{cases}\nonumber
 \end{equation}
The field $\chi$ is related to the Higgs field in the unitary gauge ($H=(0, h/\sqrt{2})^T$) in a well-defined manner\footnote{Although Eq.~\eqref{connection2} can be exactly integrated  \cite{GarciaBellido:2008ab}, the resulting expression is not very enlightening. For the purposes of this paper, it will be enough to consider just the limiting cases $h\ll M_P/\xi$ and $h\gg M_P/\xi$ in which
\begin{equation}\label{connection2}
\chi\simeq \begin{cases} 
 h\,, & h\ll \frac{M_P}{\xi}\,,\\
\sqrt{\frac{3}{2}} M_P \log \Omega^2(h)\,, & h\gg \frac{M_P}{\xi}\,. 
   \end{cases}\nonumber
\end{equation}
}
\begin{equation}\label{connection}
\frac{d\chi}{d h}=\sqrt{\frac{\Omega^2+6\xi^2h^2/M_P^2}{\Omega^4}}\,, \,\,\,\,\,\,\Omega^2 = 1+\frac{\xi h^2}{M_P^2}\,.
\end{equation}
Given the hierarchy between the electroweak and the Planck scales and the restriction $\xi\ll M_P^2/v^2_{EW}$, we can safely approximate $1+\xi v_{EW}^2/M_P^2\approx 1$ in Eq.~\eqref{potentialE1} and consider the simplified potential\footnote{The difference between the electroweak scale $v_{EW}$ and the scale $M_P/\xi$ at which the effective potential is significantly modified allows us to identify, for all practical purposes, $v_{EW}$ and $\chi=0$.}
 \begin{equation}\label{Vtree}
 V(\chi)\simeq\frac{\lambda M_P^4}{4\xi^2}\left(1-e^{-\sqrt{2/3}\,\chi/M_P}\right)^2\
 \end{equation}
for $\chi\gg \frac{M_P}{\xi}$. 
The determination of the associated inflationary observables follows the usual slow-roll approach. Taking into account the relation   between the number of e-folds $N$ and the field value at horizon exit, 
  \begin{equation}\label{efolds}
N=\frac{1}{M_P}\int_{\chi_{\rm end}}^{\chi_*} \frac{d \chi}{\sqrt{2\epsilon}}\simeq \frac{3}{4} e^{\sqrt{2/3}\chi_*/M_P}
 \end{equation}
 we can translate the  normalization of scalar perturbations at large scales [$V/\epsilon=24\pi^2\Delta_{\cal R}^2 M_P^4 \simeq (0.0276\, M_P)^4$] into a constraint 
on the ratio of couplings determining the amplitude of the inflationary potential~\eqref{Vtree}, namely
\begin{equation}\label{lxi}
\frac{ \lambda}{\xi^2}\simeq 4\times 10^{-11}\,.
\end{equation}
Equation \eqref{efolds} also allows us to compute the slow-roll parameters at horizon exit
 \begin{equation}
 \epsilon_*\simeq \frac{3}{4N^2}\,,\hspace{5mm} \eta_*=-\frac{1}{N}
 \end{equation}
 and with them the inflationary observables
 \begin{equation}
 n_s=1+2\eta-6\epsilon \simeq 1-\frac{2}{N}\,,\hspace{5mm} r=16\epsilon_*\simeq \frac{12}{N^2}\,.
 \end{equation}
The precise number of e-folds to be inserted in the previous expressions depends on the duration of the reheating stage. The efficiency of the reheating process computed in Refs.~\cite{GarciaBellido:2008ab,Bezrukov:2008ut} sets $N\simeq 59$.

 At the classical level, the model predicts a Gaussian spectrum of primordial fluctuations with a universal spectral index for scalar perturbations ($n_s \simeq0.966$) and a small tensor-to-scalar ratio ($r\simeq0.0034$) \cite{Bezrukov:2007ep}.
 No relation between particle physics and cosmological parameters shows up. The variation  of the self-coupling $\lambda$, associated with a change of the Higgs boson mass, can be always compensated by the change of the (\textit{a priori} unknown) nonminimal coupling $\xi$ (see~Eq.~\eqref{lxi}). For $\lambda\sim {\cal O}(1)$, $\xi$ is required to be rather large, $\xi \sim {\cal O}(10^4)$, but still significantly smaller than the value giving rise to noticeable effects in low-energy experiments \cite{Bezrukov:2007ep}. 
Although this large value could be considered ``unnatural'', it is commensurable to the values obtained when considering the different hierarchies already present in the Standard Model.  Rather large ratios such as $m_t/m_u\simeq 10^5$ seem to be  unnatural  but they are certainly realized in nature. Moreover, it is possible to construct embeddings of Higgs inflation in which an apparently unnatural nonminimal coupling $\xi$ appears as the low-energy remnant of a ``natural'' theory containing parameters of order one and different energy scales \cite{Barbon:2015fla}.
 
The link between the SM parameters and cosmological observations appears when quantum effects are taken into account. The inclusion of quantum corrections is a nontrivial task. When written in the Einstein frame, the Lagrangian \eqref{lagr} is essentially nonpolynomial and therefore nonrenormalizable.  This immediately  poses a number of questions: 
\begin{enumerate}[A.]
\item 
``What is the sensitivity of Higgs inflation to the higher-dimensional operators that should be included into the analysis? Which is the proper ultraviolet cutoff?'' 
\item ``Can we do reliable computations of radiative corrections in a nonrenormalizable theory?''  
\item ``What is the running of the SM couplings to the inflationary scale where the Higgs inflation potential should be computed? What is the relation between low- and high- energy parameters?''  
\end{enumerate}

In the absence of an ultraviolet (UV) completion for the SM nonminimally coupled gravity, the answer to these questions can be only based on the \emph{self-consistency} of the procedure. This was indeed the attitude taken in Ref.~\cite{Bezrukov:2010jz}, where the effective field theory for Higgs inflation was formulated (some further developments can be also found in Ref.~\cite{Bezrukov:2014bra}). In what follows we summarize the main assumptions and results of this approach and provide answers to the questions A, B and C.

\subsection{Sensitivity to higher-dimensional operators}\label{subset:A}

The naive dimensional analysis stemming from the standard effective field theory approach leads to the generic conclusion that the inflationary predictions and even the very existence of an inflationary dynamics are very sensitive to the UV completion of the low-energy theory (for a recent discussion see~Ref.~\cite{Burgess:2014lza}). Quartically and quadratically divergent loops lead to the generation of a constant term (a cosmological constant) and a quadratic term in the scalar field whose magnitude is determined by the scale of new physics.  The flatness of the effective inflationary potential at the large field values needed for inflation is generically spoiled by these powerlike corrections. This is nothing else than the ``inflationary''  manifestation of the celebrated cosmological constant and hierarchy problems which permeate all beyond-Standard-Model computations and that remain without any convincing solution. We will not try to provide any further input on these complicated problems. Instead, we will follow the standard logic of the effective field approach: we will add to the SM Lagrangian different higher-order operators without addressing the tuning of the cosmological constant and the Higgs mass to their observed values. The structure of these operators will be constrained by several self-consistent hypothesis concerning the symmetries of the UV completion, which we will specify in detail below.

Let us denote by $\Lambda$ the suppression scale of the higher-dimensional operators to be added to the SM.\footnote{For concreteness, the discussion in this section is based on the Jordan frame Lagrangian \eqref{lagr}.}  The value of $\Lambda$ is \textit{a priori} unknown and depends on the different thresholds (masses of new particles) that were integrated out to get the low-energy effective field theory. In principle, it could be as large as the Planck mass $M_P$, where gravitational interactions become important for sure. In that case, the effect of higher-dimensional operators such as $h^6/M_P^2$ would be numerically small for sufficiently large $\xi$.\footnote{For typical inflationary field values $h \propto M_P/\sqrt{\xi}$, the correction to inflationary energy density is of order $\delta V_\text{infl}/V_\text{infl} \sim 1/(\xi \lambda)$.}

Although quite natural, the identification of the cutoff scale with the Planck mass may turn out to be theoretically inconsistent since other processes can break tree-level unitarity at lower energies \cite{Burgess:2009ea,Barbon:2009ya,Burgess:2010zq,Bezrukov:2010jz}.  A self-consistent approach is to define the parameter $\Lambda$ from the theory itself by considering all the possible reactions between the SM constituents.  

The energy scale signaling the breaking of tree-level unitarity  in particle collisions depends on the expectation value of the background field $h$. At small field values ($h \lesssim M_P/\xi$), the cutoffs associated to the different interactions agree with the result of the naive computation performed around the  electroweak vacuum, $\Lambda(h) \simeq M_P/\xi$. At large field values ($h \gtrsim M_P/\sqrt{\xi}$), the suppression scale depends on the particular scattering process considered.  The suppression of graviton-graviton interactions is particularly strong and coincides with the dynamical Planck scale $\Lambda^2(h) \simeq \xi_h h^2$. The lowest cutoff dictated by the theory appears in the gauge sector and grows linearly with the field, $\Lambda(h)\sim h$ 

All the relevant scales involved in the inflationary and postinflationary evolution of the Universe
 (i.e.\ the Hubble rate and the reheating temperature of the Universe) are parametrically smaller than the previous cutoffs. As a consequence of this, the weak coupling approximation remains valid and  the cosmological predictions of the Higgs inflation are stable against the addition of higher-dimensional operators introduced along the lines of the standard effective field theory reasoning \cite{Bezrukov:2010jz} (see also Ref.~\cite{Ferrara:2010in}).  

\subsection{Reliability of the computation of radiative corrections} \label{subset:B}

Since the SM (itself renormalizable) is coupled to  gravity, the resulting theory is clearly nonrenormalizable. According to the general rules of quantum field theory, a sensible computation of radiative corrections requires the addition of an infinite number of counterterms together with the choice of a subtraction scheme.
 The structure of the counterterms will be similar to that of the higher-dimensional operators discussed above, but with their precise form fixed by the requirement  of removing the divergencies in the loop diagrams stemming from the initial Lagrangian.

Because of  nonrenormalizability, the result of the computations {\em does depend} on the subtraction scheme, the choice of which would correspond to a specific UV completion of the theory (see~Sec.~3.2 of Ref.~\cite{Bezrukov:2010jz} for details). 
Two ingredients of a potential UV completion of the Standard Model + gravity were conjectured in Refs.~\cite{Shaposhnikov:2008xb,Shaposhnikov:2008xi}, namely i) an exact, but spontaneously broken, quantum scale-invariance and ii) the absence of particles with masses larger than the electroweak scale. In this framework, physical scales appear as the consequence of the spontaneous breaking of scale invariance and are proportional to the vacuum expectation value of an extra scalar field---the dilaton. As shown in Refs.~\cite{Armillis:2013wya,Gretsch:2013ooa}, it is possible to remove the divergencies in such a way that the scale symmetry  (or the conformal symmetry, if gravity is not included) remains intact at all orders of perturbation theory.  The bare Higgs mass is then protected from large radiative corrections and the dilatational anomaly is absent (see~Ref.~\cite{Shaposhnikov:2008xi}). The  price to pay is the lack of  renormalizability \cite{Shaposhnikov:2009nk}, which does not seem to be a strong requirement given the fact that gravity itself is not renormalizable. The embedding of Higgs inflation in this type of scale-invariant framework and its self-consistency was presented in Refs.~\cite{GarciaBellido:2011de,Bezrukov:2012hx}.

Unfortunately, we do not know if an UV completion of the SM+gravity with the previous properties exists beyond perturbation theory or whether it is really realized in nature.  In what follows, we will simply assume this to be the case.

The requirement of maintaining the classical symmetries\footnote{Namely, the approximate scale invariance in the Jordan frame at large field values and the asymptotic shift symmetry for the canonically normalized scalar field in the Einstein frame.} in the quantized version of the theory  (or equivalently, the assumption of a scale-invariant UV completion) fixes the functional form of the counterterms. For a generic renormalization procedure satisfying this requirement (based, for example, on cutoff \cite{Bezrukov:2010jz} or lattice \cite{Shaposhnikov:2008ar} regularizations), the Higgs potential in the Einstein frame remains exponentially flat at large field values ($h\gtrsim M_P/\sqrt{\xi}$) and coincides with that of the Standard Model at low ones ($h\lesssim M_P/\xi$). What happens in the intermediate region ($M_P/\xi < h < M_P/\sqrt{\xi}$) depends on the renormalization procedure and reflects our  lack of understanding of the UV completion\footnote{The existence of threshold effects associated to higher-dimensional operators was also considered in Ref.~\cite{Burgess:2014lza}.}. The relation between the low-energy parameters of the SM (such as the Higgs boson mass  or the top Yukawa coupling) and the inflationary observables is generically lost (see~Ref.~\cite{Bezrukov:2010jz} and Sec.~\ref{subset:C}). However, if the dynamical evolution of the Higgs field, starting from initial chaotic inflationary conditions  $(h\gtrsim M_P/\sqrt{\xi})$, is able to bring the system to the SM vacuum  ($h\simeq 250$ GeV), the idea and predictions of Higgs inflation remain in force. In particular, \textit{Higgs inflation with a metastable electroweak vacuum can be possible}  (see~Sec.~\ref{sec:infl}). 

A subtraction scheme that fits well with scale invariance and the assumption of not having new heavy particles between the electroweak and the Planck scale is dimensional regularization, since it effectively ignores powerlike divergences and thus minimizes all the uncertainties that can be inferred by renormalization. The standard procedure in this prescription is to compute the one loop effective action from the tree-level Lagrangian density (\ref{lagr}), expand it in Laurent series in $\epsilon=(4-D)/2$ (with $D$ the fractional dimension of space-time), and add to it the necessary operators $O_n$ with coefficients $A_n/\epsilon +B_n$ to remove the divergences. The process can be extended recursively to higher-order loops. While the coefficients $A_n$ are fixed by the structure of divergences, the coefficients $B_n$ are arbitrary. This creates the first source  of uncertainties \cite{Bezrukov:2010jz}. 

A second source of uncertainties is associated with the choice of the normalization point $\mu$, which, in dimensional regularization, corrects the mismatch in the mass dimension of the coupling constants.  In renormalizable field theories, $\mu$ is arbitrary and space-time independent. Although we could certainly maintain this prescription for our nonrenormalizable field theory nothing prevents us from modifying it and allow $\mu$ to be field dependent. ``Field independency'' is indeed not a well-defined concept in a scalar-tensor theory like the one at hand and it must be associated with a particular frame \cite{Flanagan:2004bz}. As summarized in the following table, a field-independent renormalization point $\mu$ in the Jordan frame leads to a field-dependent renormalization point $\tilde \mu=\mu/\Omega$ in the Einstein frame and vice versa \cite{Bezrukov:2009db}
\begin{center}
  \begin{tabular}{l|c|c}
                   & I \cite{Bezrukov:2007ep}& II  \cite{Barvinsky:2008ia, DeSimone:2008ei,Barvinsky:2009fy,Barvinsky:2009ii}
    \\\hline
    Jordan frame   & $\mu^2 \propto M_P^2+\xi h^2$          &$\tilde \mu^2\propto  M_P^2$
    \\\hline 
    Einstein frame & $\mu^2\propto M_P^2$                  &
                                    $ \tilde\mu^2\propto \frac{M_P^4}{M_P^2+\xi h^2}$
                                                    
  \end{tabular}
\end{center}
Choosing a particular prescription is equivalent to making some assumptions about the ultraviolet completion of the model. 
In this work, we will consider the prescription I, leaving the analysis 
with prescription II for future investigations. The reason for this choice is based on the following physical considerations. On the one hand, it allows us to maintain the quantum version of the theory as asymptotically scale-invariant at large values of the Higgs field ($h \gg M_P/\sqrt{\xi}$). On the other hand, it becomes the standard (space-time-independent) prescription of renormalizable field theories when written in the Einstein frame.

It may seem that an arbitrariness of the coefficients $B_n$ results on the loss of predictivity of the Higgs inflation. However, as it was shown in Ref.~\cite{Bezrukov:2010jz},  the finite parts of the counterterm do not change the asymptotic behavior of the scalar potential. The predictions of Higgs inflation fall into two categories:
\begin{enumerate}[1.]
\item \textit{Universal/Noncritical regime}:  For a large fraction of the parameter space, the renormalization group enhanced (RGE) potential maintains the shape and predictions of the tree-level potential. As \eqref{Vtree}, the RGE potential depends on $\lambda$ and $\xi$ only through  the combination $\lambda/\xi^2$.  Taking into account the COBE normalization and the value of the Higgs self-coupling at the inflationary scale, we can unequivocally fix the value of $\xi$, which turns out to be of order $\xi\sim{\cal O}(10^3)$.  Although the nonminimal coupling is still rather large, the difference between the scale below which the SM is valid without modifications ($M_P/\xi \sim 3\times 10^{15}$~GeV) and the scale at which inflation takes place ($M_P/\sqrt{\xi} \sim 10^{17}$~GeV) is relatively small. 
\item \textit{Critical regime}:  For very specific values of the SM parameters, the second derivative of the RGE potential becomes equal to zero at some intermediate field value between the beginning and the end of inflation. The first derivative is extremely small in the same point (but nonvanishing). This gives rise to a nonmonotonic behavior of the slow-roll parameter $\epsilon$ and opens the possibility of obtaining a sizable tensor-to-scalar ratio $r$, whose precise value strongly depends on $\xi$ and on the Higgs and top Yukawa couplings at the inflationary scale \cite{Bezrukov:2014bra,Hamada:2014iga}. The nonminimal coupling $\xi$ is generically rather small ($\xi\sim10$) and  the model does not require the inclusion of a cutoff scale significantly below the Planck scale (see~Sec.~\ref{subset:A}).
\end{enumerate}

\subsection{Relation between low- and high-energy parameters} \label{subset:C}

An analysis of Higgs inflation and its connection with low-energy observables  has been presented in Refs \cite{Bezrukov:2010jz,Bezrukov:2014bra}.  The self-consistent  set of assumptions about nonrenormalizable contributions to the action of the theory is formulated as follows:
\begin{enumerate}[i.]
\item We will only add the higher-dimensional operators that are generated via radiative corrections by the Lagrangian of the SM nonminimally coupled to gravity. In other words, only a subclass of the operators described in Sec.~\ref{subset:B} will be considered.
\item The coefficients  $B_n$  are small ($B_n \ll 1$) and have the same hierarchy as the loop corrections producing them, i.e. the coefficients in front of the operators coming from two-loop diagrams are much smaller than those coming from one-loop diagrams, etc.
\item  The renormalization scale is defined according to prescription I. This is equivalent to the requirement of scale invariance of the 
UV complete theory at large values of the Higgs field background.
\end{enumerate}
None of these assumptions are essential for the scenario presented in this paper, but they are needed to provide a (partially) controllable link between the low-energy and high-energy parameters of the model.

In this framework, the relation between {\em low-energy parameters}, such as the Higgs mass or the top quark Yukawa coupling $y_t$,  and the high-energy parameters fixing the form of the effective potential in the inflationary region does depend on the unknown coefficients $B_n$, which should be fixed by an eventual ultraviolet completion.

For $h\lesssim M_P/\xi$ the contribution of the higher-order operators $O_n$ defined in Sec.~\ref{subset:B} is suppressed. The theory is effectively renormalizable and the running of the coupling constants is governed by the usual SM renormalization group (RG) equations. In the inflationary region, i.e for $h\gtrsim M_P/\sqrt{\xi}$,  the radial component of the Higgs field is effectively frozen and the evolution of the coupling constants is determined by the renormalization group equations of the chiral Standard Model  \cite{Longhitano:1980iz} .  In both regimes, the coefficients $B_n$ do not play any role, because of the specific asymptotics of the operators $O_n$ as  functions of the Higgs field. However, in the transition region around $h \simeq M_P/\xi$, the coupling constants  change rapidly (very roughly, making a jump) by an amount proportional to the coefficients $B_n$ in front of the corresponding operators $O_n$.  

In the following section, we will compute the effective potential and will determine the magnitudes and signs of the coefficients $B_n$ giving rise to Higgs inflation in the case of a metastable electroweak vacuum.

\section{High-energy versus low-energy parameters of the Standard Model}\label{sec:hig}

In this section we set the formalism to obtain the effective action in the whole region between the SM-like regime ($\chi\ll M_P/\xi$) and the inflationary regime ($\chi>M_P/\sqrt{\xi}$). The approach described in this section closely follows the one outlined in Sec.~3 of Ref.~\cite{Bezrukov:2010jz}. Throughout this section we work exclusively in the Einstein frame and neglect the higher-order corrections in slow roll, i.e.\ we assume that all the important effects are described by the corrections to the effective potential.  

Following the discussion of Sec.~\ref{subset:B}, we will compute the effective action in dimensional regularization, where all powerlike divergences are systematically ignored. We will concentrate on the most important contribution: the one associated with the Higgs and top quark interactions\footnote{All other SM particles may be added and treated analogously.}. 
The relevant piece of the Einstein-frame Lagrangian density is given by\footnote{For illustrative purposes we will neglect the $SU(2)$ structure of the Higgs doublet and the colors of the top quark. These are not important for the derivation of Eqs. \eqref{lambdajump} and \eqref{ytjump}. } 
\begin{equation}
  \label{Ltree}
  \cL = \frac{(\partial\chi)^2}{2} - \frac{\lambda}{4} F^4(\chi)
  + i\bar\psi_t\slashed\partial\psi_t + \frac{y_t}{\sqrt{2}} F(\chi) \bar\psi_t\psi_t\,.
\end{equation}
The function $F(\chi) \equiv h(\chi)/\Omega(\chi)$ coincides with the Higgs field at low energies and encodes all the nonlinearities associated to the nonminimal coupling to gravity in the large field regime
\begin{equation}
F(\chi) \approx
  \left\{
    \begin{array}{l@{,\ }l}
      \chi & \chi<\frac{M_P}{\xi}\\ 
      \frac{M_P}{\sqrt{\xi}}\left(1-\e^{-\sqrt{2/3}\chi/M_P}\right)^{1/2}
           & \chi>\frac{M_P}{\xi}
    \end{array}
  \right\}\,.
\end{equation}
\subsection{Higgs coupling}

Let us start by computing the effective potential for (\ref{Ltree}) at one loop. We get the following two vacuum diagrams
\begin{eqnarray}
  \label{1loopVscalar0}
  \tikz[baseline=-0.5ex]{\draw[higgsline] (0,0) circle (1.4em);}
  & =& \frac{1}{2}\Tr\ln\left[\Box-\left(\frac{\lambda}{4}(F^4)''\right)^2\right]\,, \\
   \label{1loopVfermion0}
  \tikz[baseline=-0.5ex]{\draw[fermionline] (0,-1.5em) arc (270:-90:1.4em);}
  & =& -\Tr\ln\left[i\slashed\partial+y_tF\right] \,,
  \end{eqnarray}
  whose evaluation, using the standard techniques, gives 
\begin{equation}
    \label{1loopVscalar}
 \hspace{-2mm}   \tikz[baseline=-0.5ex]{\draw[higgsline] (0,0) circle (1.4em);} = \frac{9\lambda^2}{64\pi^2}\left(
   \frac{2}{\bar \epsilon}
    -\ln\frac{\lambda(F^4)''}{4\mu^2}+\frac{3}{2}
  \right) \left(F'^2+\frac{1}{3}F''F\right)^2F^4, \nonumber
    \end{equation}
\begin{equation}
 \hspace{-1mm} \label{1loopVfermion}
  \tikz[baseline=-0.5ex]{\draw[fermionline] (0,-1.5em) arc (270:-90:1.4em);}
  = 
   -\frac{y_t^4}{64\pi^2}\left(
    \frac{2}{\bar\epsilon}
    -\ln\frac{y_t^2F^2}{2\mu^2}+\frac{3}{2}
  \right) F^4\,.
\end{equation}
Here $2/\bar\epsilon$ stands for the combination $2/\epsilon-\gamma+\ln4\pi$ and the primes denote derivatives with respect to $\chi$. The divergencies in the loop diagrams \eqref{1loopVfermion} are eliminated, as usual, by adding counterterms with the definite coefficients in $1/\bar\epsilon$ and arbitrary finite parts $\delta\lambda_1$ and $\delta\lambda_2$
\begin{eqnarray}\label{counterL}
  \label{Lct}
  \delta\cL_\text{ct} &=& 
  \left(
    -\frac{2}{\bar\epsilon}
    \frac{9\lambda^2}{64\pi^2}+\delta\lambda_1
  \right) \left(F'^2+\frac{1}{3}F''F^2\right)^2F^4 \nonumber \\
  &+& \left(
    \frac{2}{\bar\epsilon}
    \frac{y_t^4}{64\pi^2}-\delta\lambda_2
  \right) F^4.
\end{eqnarray}

The effective potential is the sum of (\ref{Ltree}), (\ref{1loopVfermion}), and (\ref{Lct}), with the poles in $1/\bar\epsilon$ canceling between the counterterms and the one-loop contributions.  The structure of the counterterm involving $\delta\lambda_2$ coincides with that of the tree-level potential \eqref{Vtree}. This allows us to eliminate the constant $\delta\lambda_2$ by incorporating it into the definition of $\lambda$. The constant $\delta\lambda_1$, on the contrary, cannot be reabsorbed. It should be promoted to a new independent coupling constant with its own RG equation. The RG equations are obtained by requiring the Lagrangian density $\cL+\delta\cL$ to be independent of $\mu$.  Since we are dealing with a nonrenormalizable theory, the set of RG equations  is not closed. However, as shown in  Appendix \ref{sec:two loop}, the (infinite) system of equations can be truncated due to point (ii) in Sec.~\ref{subset:C}.

The value of $\lambda$ at the inflationary scale ($\chi \sim M_P$) depends on the counterterm (\ref{Lct}). For small field values ($F(\chi)\sim\chi\ll M_P/\xi$), the conformal factor $\Omega(\chi)$ equals to one and the theory becomes indistinguishable from the renormalizable SM.  In that case, the first term in Eq.~\eqref {Lct} turns into  a simple $\delta\lambda_1\chi^4/4$ term, which allows us to reabsorb the constant $\delta\lambda_1$ into the definition of $\lambda$. At large field values  [$F(\chi)\sim M_P/\sqrt{\xi}$, $\chi\gtrsim M_P$], the counterterm is exponentially suppressed ($\sim\delta\lambda \frac{M_P^4}{\xi^4}\e^{-4\chi/\sqrt{6}M_p}$) and the previously absorbed contribution to $\lambda$ effectively disappears. Neglecting the running of $\delta\lambda_1$ between the scales $\mu\sim M_P/\xi$ and $M_P/\sqrt{\xi}$, we can imitate this effect by a change
\begin{equation}
  \label{lambdajump}
  \lambda(\mu) \to \lambda(\mu)
  + \delta\lambda \left[
    \left(F'^2+\frac{1}{3}F''F\right)^2-1
    \right],
\end{equation}
where $\lambda(\mu)$ is evaluated using the SM RG equations. Since the effective potential is $\mu$-independent, we can choose the most convenient value of $\mu$. In order to minimize the logarithms in the one-loop contributions\footnote{If the logarithms were large, one should add higher loop contributions.},  we will take
\begin{equation}
  \label{muchoice}
  \mu^2 = \alpha m_t(\chi) = \alpha y_t F(\chi)\,,
\end{equation}
with $\alpha$ a constant of order one. This choice of $\mu$ is consistent with prescription I in Sec.~\ref{subset:B}.

\subsection{Top Yukawa coupling}
The effect described above applies also to the Yukawa coupling.  To see this, consider the propagation of the top quark in the background $\chi$ 
\begin{equation}
  \tikz[baseline=-0.5ex] {
    \draw[fermionline] (-3em,0) -- (3em,0);
    \draw[higgsline] (-1.5em,0) arc (180:0:1.5em);
    \draw[fill] (-1.5em,0) circle (0.2em) node[below] {$yF'$};
    \draw[fill] (1.5em,0) circle (0.2em) node[below] {$yF'$};
  }
  +
  \tikz[baseline=-0.5ex] {
    \draw[fermionline] (-3em,0) -- (0,0);
    \draw[fermionline] (0,0) -- (3em,0);
    \draw[higgsline] (0,0) arc (-90:270:1.5em);
    \draw[fill] (0,0) circle (0.2em) node[below] {$yF''$};
  }
\end{equation}
Canceling divergencies in these diagrams requires the counterterms of the form
\begin{eqnarray}\label{counteryt}
  \delta\cL_\text{ct} &\sim&
  \left(\#\frac{y_t^3}{\bar\epsilon}+\delta y_{t1}\right) F'^2F \bar\psi\psi \nonumber \\
  &+&
  \left(\#\frac{y_t\lambda}{\bar\epsilon}+\delta y_{t2}\right) F''(F^4)''\bar\psi\psi  
  ,
\end{eqnarray}
where $y_tF'$ appears from the vertices with one Higgs field, $y_tF''$ from vertices with two Higgs fields, $y_tF$ from the mass of the top quark in the propagator and $\lambda(F^4)''$ from the mass of the scalar propagator in the bubble.
The term with $\delta y_{t1}$ has similar properties to that with $\delta\lambda_1$.  In the limit of small $\chi$ it goes as $\chi\bar\psi\psi$, and it can be reabsorbed into the definition of the Yukawa coupling $y_t$ (as in the SM). For large $\chi$, the counterterm vanishes and the contribution $\delta y_t$ into $y_t$ disappears. As before, we neglect the running of $\delta y_{t1}$ between $M_P/\xi$ and $M_P/\sqrt{\xi}$ and parametrize this effect by an effective change
\begin{equation}
  \label{ytjump}
  y_t(\mu) \to y_t(\mu)+ \delta y_t\left[
    F'^2-1
  \right]\,,
\end{equation}
with $\mu$ given by Eq.~\eqref{muchoice}.

\section{Higgs inflation with metastable vacuum}\label{sec:infl}

If the ``jumps''  $\delta\lambda$ and $\delta y_t$ are much smaller than the respective coupling constants at the transition scale $M_P/\xi$ [$\delta\lambda \ll \lambda(M_P/\xi)$, $\delta y_t \ll y_t(M_P/\xi)$], Higgs inflation requires the absolute stability of the vacuum and provides a clear connection between the properties of the Universe at large scales and the value of the SM Higgs and top quark masses. However, since the smallness of $\lambda$ at the inflationary scale appears as the result of a nontrivial cancellation between the fermionic and bosonic contributions, it is reasonable to think that $\delta\lambda$ can be commensurable to $\lambda$. In that case, the jumps of the coupling constants open the possibility of having Higgs inflation even in the case of a metastable vacuum by converting a negative scalar self-coupling below $M_P/\xi$ into a positive coupling above that scale. Some illustrative values of the parameters needed to restore noncritical Higgs inflation beyond an instability scale $\mu_0$ are presented in Table \ref{table1}. 

\begin{table}
\setlength{\tabcolsep}{6pt}
\begin{center}
\begin{tabular}{SSSSS} \hline
\hline \noalign{\smallskip} 
 {$m_h=125.5$ GeV}&{$\xi=1500$} &\hspace{5mm}&{$\delta y_t\simeq 0.025$}  \\   \noalign{\smallskip} \hline \hline \noalign{\smallskip} 
 {$m_t$}&  {$\mu_0$ (GeV)}&  \hspace{5mm}& {$\delta\lambda$}   \\ \noalign{\smallskip} \hline  
  \noalign{\smallskip}
  172.0 &{$\sim 2 \times 10^{12}$}&\hspace{5mm}&  -0.008  \\
 173.1 &{$\sim 2 \times 10^{10}$}&\hspace{5mm}&  -0.015  \\
 174.0 &{$\sim 2 \times 10^{9}$}&\hspace{5mm}& -0.022\\
175.0 &{$\sim 3 \times 10^{8}$} &\hspace{5mm}& -0.029 \\ \hline\hline 
\end{tabular}
\end{center}
\caption{Illustrative values of the top pole mass $m_t$ and the associated instability scale $\mu_0$ for fixed values of the Higgs mass $m_h$ and the nonminimal coupling to gravity $\xi$. The  values of $\delta \lambda$ are chosen to restore the asymptotic behavior of the potential at the inflationary scale. All choices of parameters give roughly  the same inflationary predictions.}\label{table1}
\end{table}
\begin{figure}
\centering
\includegraphics[scale=0.41]{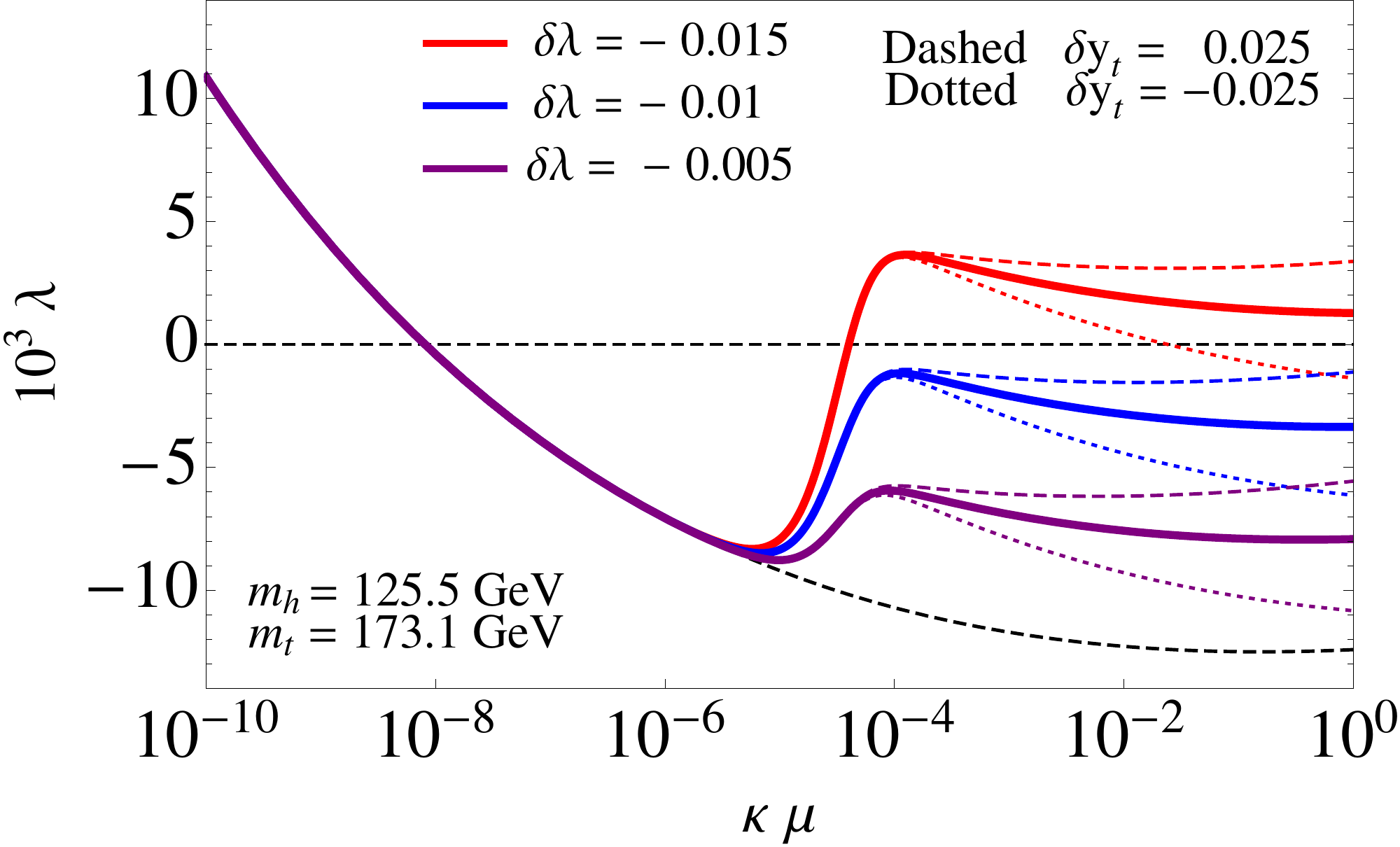}
\includegraphics[scale=0.41]{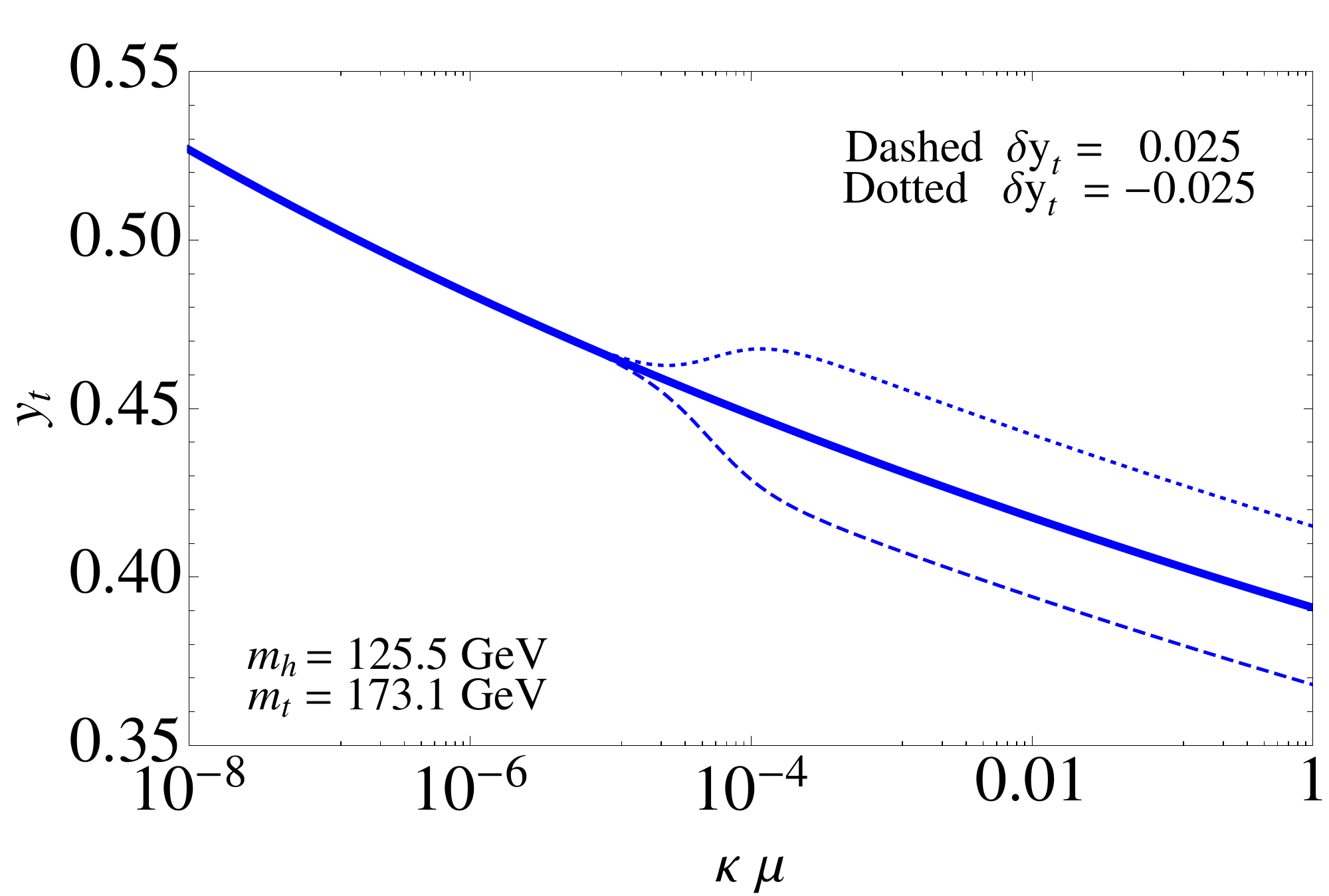}
\caption{(color online) Effect of the coefficients $\delta\lambda$ and $\delta y_t$ on the running of $\lambda$ and $y_t$ as a function of the field-dependent renormalization scale $\mu(\chi)$ in Planck units ($\kappa=M_P^{-1}$). The effect of the jumps is localized at an energy scale $\mu(\chi\sim M_P/\xi)$ and does not significantly modify the asymptotic behavior of the couplings in the low- and high- energy regions. This property ensures that the inflationary predictions of the critical and noncritical scenarios (see~Sec.~\ref{subset:B}) are maintained even in the presence of jumps.}\label{fig:jumps}
\end{figure}
The effect of the coefficients $\delta\lambda$ and $\delta y_t$ on the running of the coupling $\lambda$ and $y_t$  is summarized in Fig.~\ref{fig:jumps}. Qualitatively, the coefficient $\delta\lambda$ controls the height of the potential in the inflationary region, while the coefficient $\delta y_t$ controls the tilt. As schematically represented in Fig.~\ref{cartoon}, the (zero-temperature) effective potential has an inflationary plateau and two minima. The asymptotic shape at large field values coincides with the one that would have been obtained \textit{in the absence of jumps and with suitable values of the Higgs mass and top Yukawa couplings}\footnote{This property ensures that the inflationary predictions of the critical and noncritical scenarios (see~Sec.~re\ref{noncriticalsubset:B}) are maintained even in the presence of jumps.}. The first minimum (the shallowest and narrowest one) corresponds to the standard electroweak vacuum\footnote{Remember that at low field values ($\chi\ll M_P/\xi$) the effective potential coincides with that of the usual Standard Model minimally coupled to gravity [see~Eq.~\eqref{potentialE1}]. }. The second minimum (the deepest and widest one) is generated by the interplay between the instability of the Higgs self-coupling below $M_P/\xi$ and the jumps at that scale. As in any chaotic inflation scenario, the Higgs field will start its evolution from trans-Planckian values, will inflate the Universe and will decay into the SM particles after the exponential expansion \cite{Linde:2005ht}. At first sight, it may seem that, at the end of this set of processes, the Universe will end at the deeper and wider vacuum at $\chi\sim M_P/\xi$.
 However, this is not necessarily the case. The destiny of the Universe strongly depends on the relation between the energy stored in the Higgs field after inflation and the depth of the minimum at large field values. If the first of these is much larger than the second, the reheating of the Universe after inflation may result into a sizable modification of the effective potential, leading to the disappearance of the ``dangerous'' vacuum at large field values and the subsequent evolution of the system towards the ``safe'' electroweak vacuum. On the other hand, if the two energy scales are comparable, the Universe will end in the ``dangerous'' vacuum and will inevitably collapse \cite{Felder:2002jk}. 

\begin{figure}
\centering
\includegraphics[scale=0.41]{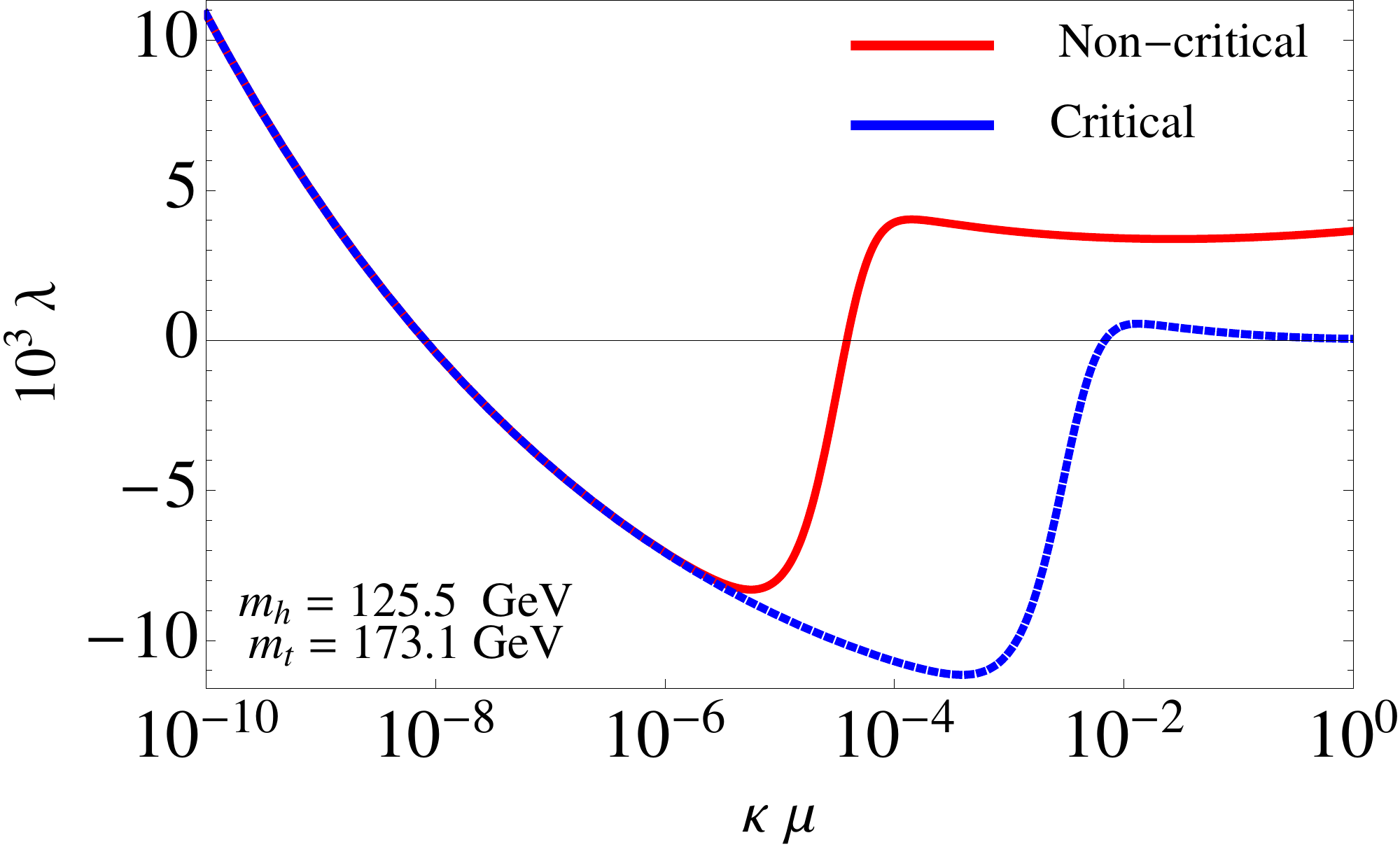}
\caption{(color online) Comparison between the running of the Higgs self-coupling $\lambda$ in the critical ($\xi=15$, $\delta y_t=0$, $\delta\lambda=-0.0133$) and noncritical  ($\xi=1500$, $\delta y_t=0.025$, $\delta\lambda=-0.015$) cases as a function of the field-dependent renormalization scale $\mu(\chi)$ in Planck units ($\kappa=M_P^{-1}$).}\label{lmu}
\end{figure}
 The depth and width of the extra minimum is effectively controlled by the nonminimal coupling $\xi$, which determines the value $\chi\sim M_P/\xi$ at which the transition from negative to positive $\lambda$ takes place (see~Fig.~\ref{lmu})\footnote{Strictly speaking, the values of the parameters for the noncritical case presented in Fig.~\ref{lmu} \textit{do not} give rise to realistic values of the inflationary observables and are included just for illustration purposes. The proper analysis of the parameter space in critical Higgs inflation is extremely subtle and we postpone it to a future work.}. At the same time, $\xi$ is the main parameter distinguishing the noncritical and critical scenarios\footnote{Note that these two regimes can be distinguished cosmologically: in the noncritical case the tensor to scalar ratio is small, while in the critical case it can be rather large.}. The dual role of $\xi$ allows us to conclude that the depth of the minimum is generically much smaller than the scale of inflation in the noncritical case and comparable to it in the critical one. Critical Higgs inflation is then expected to require the absolute stability of the vacuum. 

The following sections are devoted to quantifying the general arguments presented above. In Sec.~V, we will discuss the finite-temperature effective potential and determine the minimal temperature $T_+$ needed to restore the stability of the potential in the noncritical and critical cases. By comparing this temperature with the upper bounds on the reheating temperature $T_{RH}$ obtained in Sec.~\ref{sec:preh0}, we will demonstrate that $T_{RH}>T_+$ in noncritical case and that $T_{RH}<T_+$ in the critical one.

\vspace{0.5cm}

\section{High-temperature effective potential}\label{sec:eff}

The set of coefficients presented in Table \ref{table1} makes the Higgs self-coupling positive at large values of the Higgs field and allows for inflation. However, the zero-temperature effective potential has an extra minimum at large values of the scalar fields. In this section, we consider the change in the shape of the effective potential in the presence of a thermal plasma, like that originated  by the decay of  the inflaton into the SM particles. In particular, we will determine the minimum temperature needed to stabilize the effective potential and the temperature needed to drive the Higgs field towards the true electroweak minimum.

The one-loop finite temperature corrections can be written in  the form \cite{Linde:1978px}
\be\label{thermal}
\Delta V_T= -\frac{1}{6\pi^2}\sum_{B,F} \int_0^\infty \frac{k^4 dk}{\epsilon_k(m_{B,F})} n_{B,F}[\epsilon_k(m_{B,F})]~.
\ee
Here $n_B$ and $n_F$ are the Bose and Fermi distributions $n_{B,F}[x]=1/(e^{x/T}\mp 1)$, $\epsilon_k(m) = \sqrt{k^2 + m^2}$, $m_{B,F}$ are the masses of SM particles in the background Higgs field, and the summation is over all the SM degrees of freedom.
The most important contributions come from the top quark and the gauge bosons, with masses
\begin{align}
  m_Z^2(T) &= \frac{g_1^2(\mu_g)+g_2^2(\mu_g)}{4}F^2(\chi),\\
  m_W^2(T) &= \frac{g_2^2(\mu_g)}{4}F^2(\chi), \\
  m_t(T) &= \frac{y_t(\mu_t)}{\sqrt{2}}F(\chi) .
\end{align}
The coupling constants in the previous expressions should be taken at the relevant scale, which can be always chosen proportional to the temperature.\footnote{The coupling constants appearing in boson loops were evaluated at the scale $\mu_g=7T$. On the other hand, those appearing in fermion loops were evaluated at the scale $\mu_t=1.8T$. As shown in Ref. \cite{Kajantie:1995dw}, this choice minimizes radiative corrections. The previous two choices should be  replaced  by more complicated expressions involving $\chi$ and $T$ in the limit of low temperatures (smaller that the background field scale). This change is however irrelevant from a numerical point of view, since in that case the thermal potential \eqref{thermal} is exponentially suppressed.}
\begin{figure}
\centering
\includegraphics[width=\columnwidth]{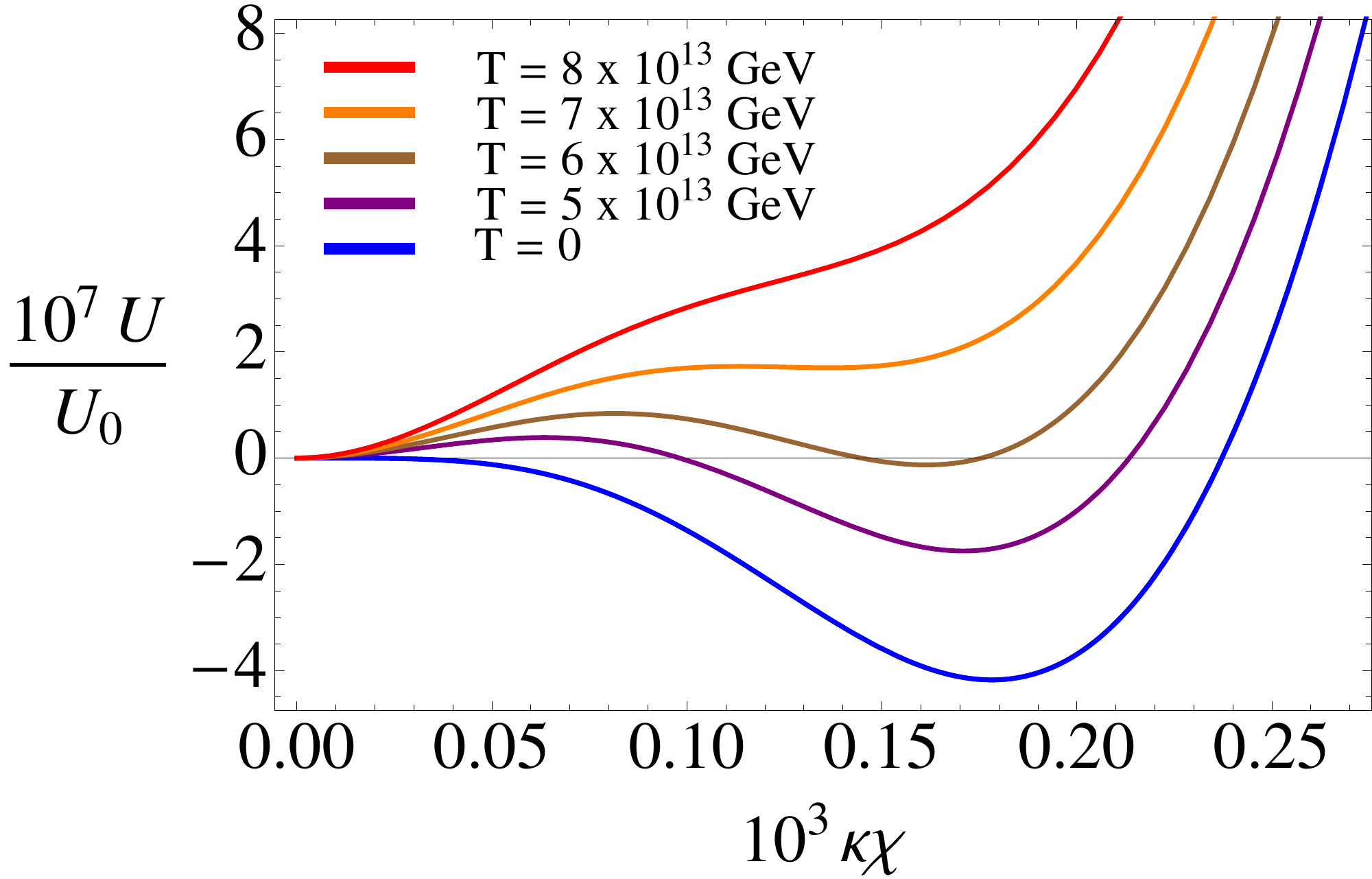}
\includegraphics[width=\columnwidth]{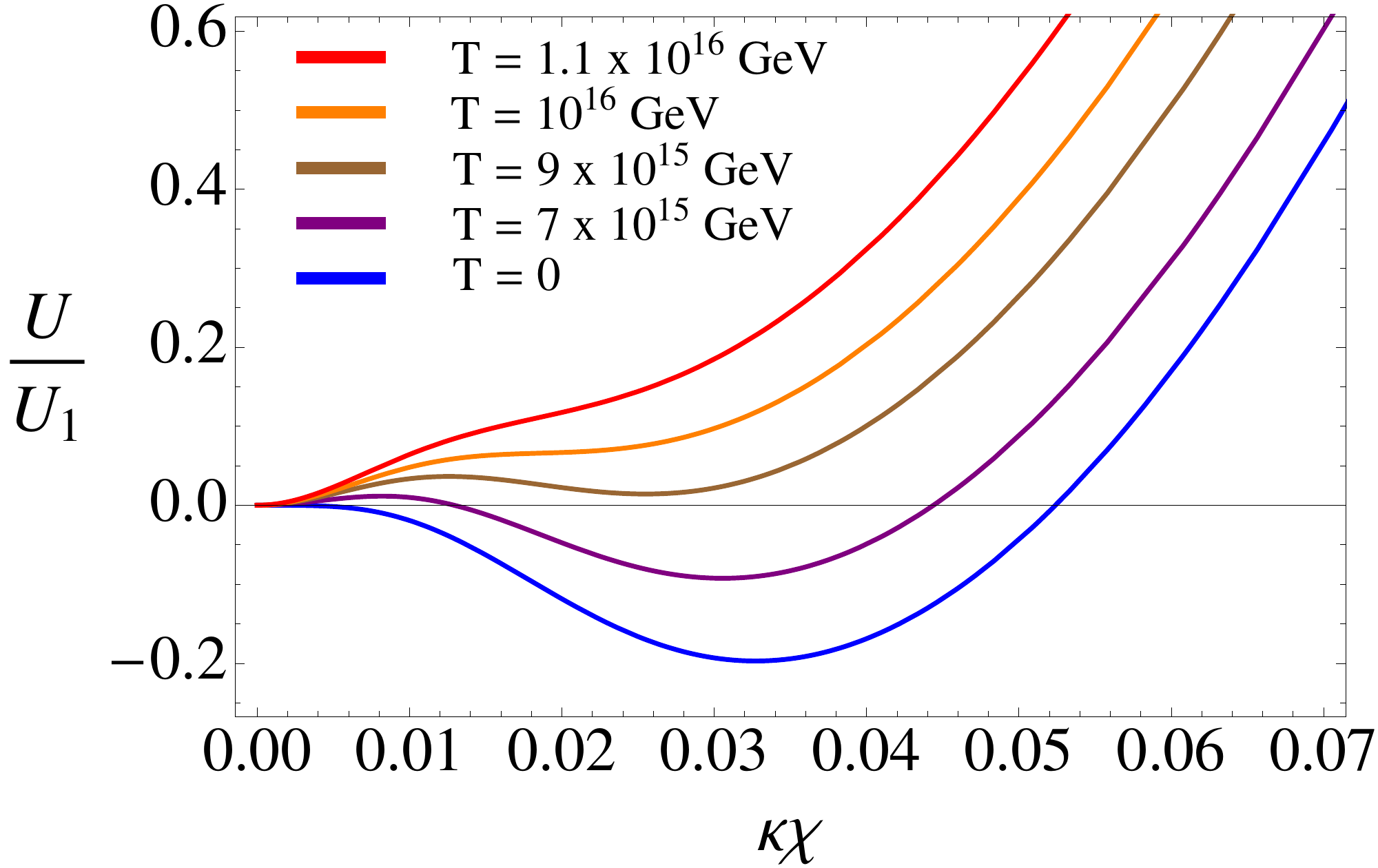}
\caption{(color online) Top: High-temperature effective potential for noncritical Higgs inflation ($m_h=125.5$\,GeV, $m_t= 173.1$\,GeV, $\xi=1500$, $\delta y_t = 0.025$, $\delta\lambda = -0.0153$). 
Bottom: High-temperature effective potential for critical Higgs inflation ($m_h=125.5$\,GeV, $m_t= 173.1$\,GeV, $\xi=15$, $\delta y_t = 0$, $\delta\lambda = -0.01325$).}
\label{fig:thpot}
\end{figure}
The thermally corrected effective potential for the noncritical/critical case is shown in the upper/lower part of Fig.~\ref{fig:thpot}. In the noncritical case, the restoration temperature and the temperature at which the minimum at high field values of the Higgs field disappears are given respectively by
\begin{equation}\label{TcritNC}
T_-^{NC}\simeq6\times10^{13}\,\,\,\text{GeV}, \hspace{10mm}T^{NC}_+\simeq7\times10^{13}\,\,\,\text{GeV}\,.
\end{equation}
 The associated temperatures in the critical case turn out to be significantly larger\footnote{Remember that the second minimum of the potential is much wider and deeper in the critical case.}
\begin{equation}\label{TcritC}
T_-^{C}\simeq9\times10^{15}\,\,\,\text{GeV}, \hspace{10mm}T^C_+\simeq10^{16}\,\,\,\text{GeV}\,.
\end{equation}
If the Universe is heated up to the temperatures $T_{RH}$ above $T_+$, the system will relax to the SM vacuum.  In the subsequent evolution of the Universe the temperature decreases and the second minimum reappears at large field values, first as a local minimum, then as the global one. However, there is always a barrier separating these two minima (not really visible on the plot due to the overall $\chi^4$ behavior at low field values). This barrier prevents the direct decay of the Fermi vacuum. 
\begin{figure}
\centering
\includegraphics[width=\columnwidth]{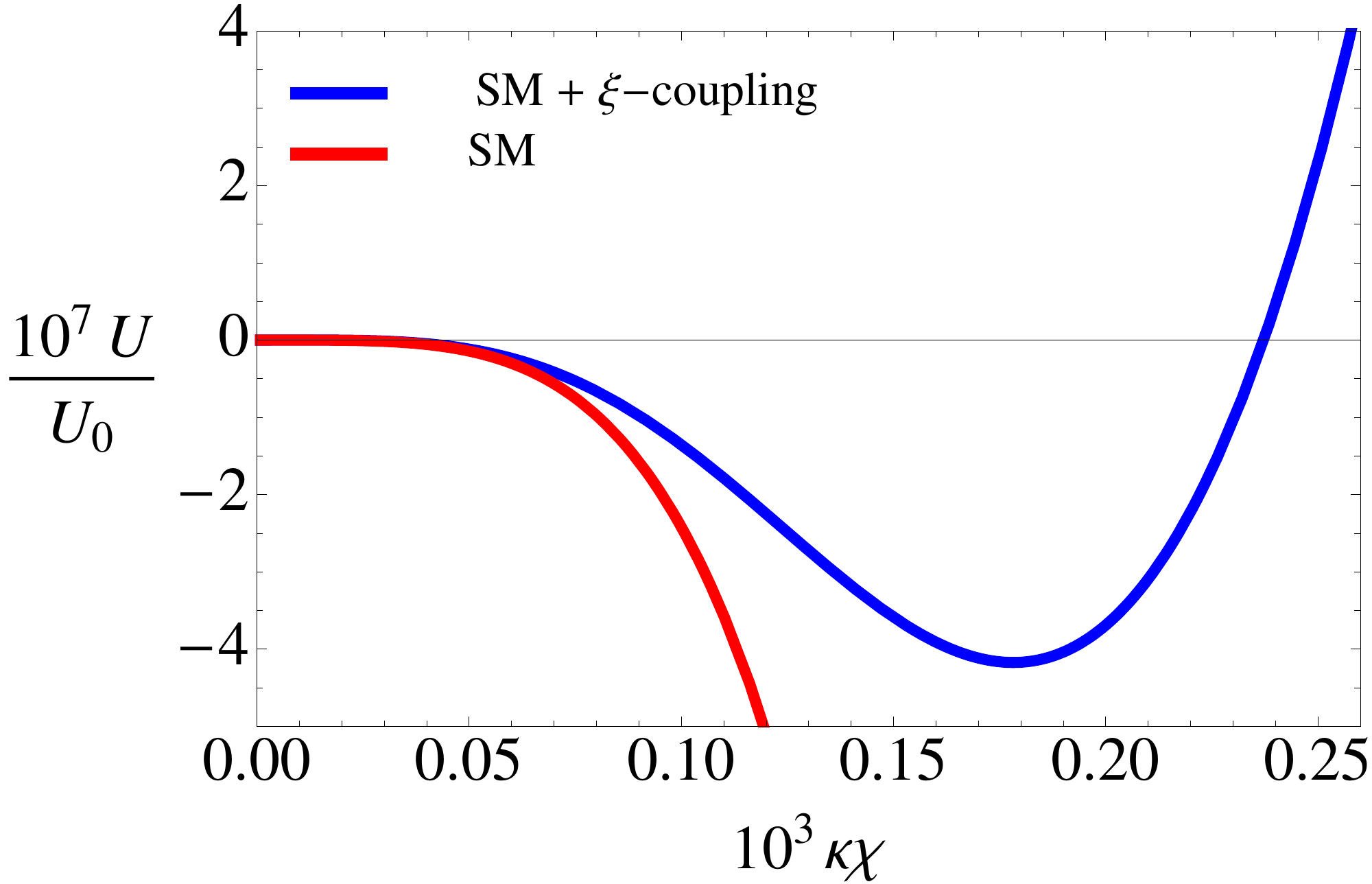}
\caption{Comparison between the effective potential for the SM and the SM nonminimally coupled to gravity ($\xi=1500$,  $\delta y_t = 0$, $\delta\lambda = -0.017$). The normalization scale $U_0=(10^{-3} M_P)^4$ coincides with that in Fig.~\ref{potfig} and $\kappa=M_P^{-1}$.  Since the potential for the nonminimally coupled case lays on top of the SM, the lifetime of the Universe in Higgs inflation is even larger than in the SM alone.}
\label{fig:thSMcomp}
\end{figure}
The decay of the SM vacuum can still happen via tunneling, but the probability of this process turns out to be rather small. The probability of the decay of the EW vacuum due to thermal fluctuations in the absence of gravity was studied in Refs.~\cite{Espinosa:2007qp,Anderson:1990aa,Arnold:1991cv,Espinosa:1995se}, with the result that this effect does not lead to the vacuum decay for the present-day values of the Higgs and top quark masses.  Since the  thermal potential is always above the SM one (see~Fig.~\ref{fig:thSMcomp}), the probability of decay in our case is even smaller than in the absence of gravity and can be safely neglected.

The aim of  the next sections is to estimate the reheating temperature $T_{RH}$ in the critical and noncritical cases. It should be noted that the effect of the symmetry restoration discussed above {\em does not require thermal equilibrium} \cite{Tkachev:1995md}. What is important is that the medium effects change the effective potential in such a way that the positive contributions to the scalar mass are generated. The effective temperature $T_*$ that can be used for an estimate of the medium effects can be defined through the typical integrals that appear in the computation of the effective potential,
\be\label{Tstar}
\frac{T_*^2}{24} \simeq \int \frac{d^3 k}{2|k|(2\pi)^3}n^{\rm noneq}_{B,F}~,
\ee
where $n^{\rm noneq}_{B,F}$ are the distributions of the particles created at preheating. The preheating temperate $T_{RH}$ determined in the next sections is generically smaller than $T_*$ and it should be then understood as a \textit{conservative} estimate of the temperature to be compared with the restoration temperature $T_+$.
\section{Preheating}\label{sec:preh0}

Determining the proper reheating temperature is a rather complicated task. It generically requires the use of numerical simulations able to deal with the highly nonlinear and nonperturbative particle production after inflation together with a detailed analysis of the thermalization stage (see Refs. \cite{
Allahverdi:2010xz,Amin:2014eta} for a review). In this section, we will simply try to provide a rough estimate of this temperature based on the following considerations:

\begin{enumerate}
\item  At the end of inflation the Higgs field oscillates around the minimum of the potential. During each semioscillation $j$, the SM fields coupled to it oscillate many times and particle creation takes place. The depletion of the Higgs condensate is dominated by the production of $W$ and $Z$ bosons\footnote{The direct production of fermions is suppressed by Pauli blocking effects.}${}^{,}$\footnote{The creation of Higgses, although tachyonic in nature, is much less efficient (see\ Appendix \ref{appendix2} for details).} and their subsequent decay into relativistic SM fermions. 

The interplay between nonperturbative particle creation and decays will be accounted for using the \textit{combined preheating} formalism \cite{GarciaBellido:2008ab,Bezrukov:2008ut}, which allows us to estimate the energy density of the different species as a function of the number of semioscillations (see~Appendix \ref{appendix3} for details and notation). The beginning of the radiation-domination era will be determined by the time at which the energy in the relativistic fermions equals the energy density in the homogeneous background field. 

\item  At the time of production, the distribution of fermions is far from thermal. To achieve equilibrium, they  must interact to redistribute their energy (kinetic equilibrium) and to adjust their number density (chemical equilibrium).  As shown in Ref.~\cite{Kurkela:2011ti} (see also Refs.~\cite{Enqvist:1990dp,Davidson:2000er,Mazumdar:2013gya}, the particular way in which this happens depends on the relative occupancy of the produced plasma with respect to a thermal distribution. To determine if we are dealing with an under- or overoccupied system, we will define an instantaneous ``radiation temperature''\footnote{This temperature is obtained by equating the total energy density of the fermions at a given time, $\rho_F(j)$, to the energy density of a thermal plasma in thermal equilibrium.}
 
\begin{equation} \label{Tformula}
 T_{r}(j) \equiv \left(\frac{30\rho_F(j)}{g_*\pi^2}  \right)^{1/4}\,,
\end{equation} 
and we will compare the number density and average energy per particle in our plasma to those of a thermal gas containing the same number degrees of freedom\footnote{The factor  $g_*$ stands for the number of degrees of freedom resulting from the decay of $W$ and $Z$ bosons. Since the top quark is not produced in these decays, we will take $g_*=(4\times3+2\times3+4\times 3\times 5)=68.25$.}, namely
\begin{eqnarray}
n_\text{th}(j)&=&\frac{3\zeta(3)}{4\pi^2}g_*T_r(j)^3\,,\label{nth}
\\ \langle E_\text{th}(j)\rangle &=&\frac{7\pi^4}{180\zeta(3)}T_r(j)\,.\label{Eth}
\end{eqnarray}
\end{enumerate}
  
\begin{figure}
\includegraphics[scale=0.39]{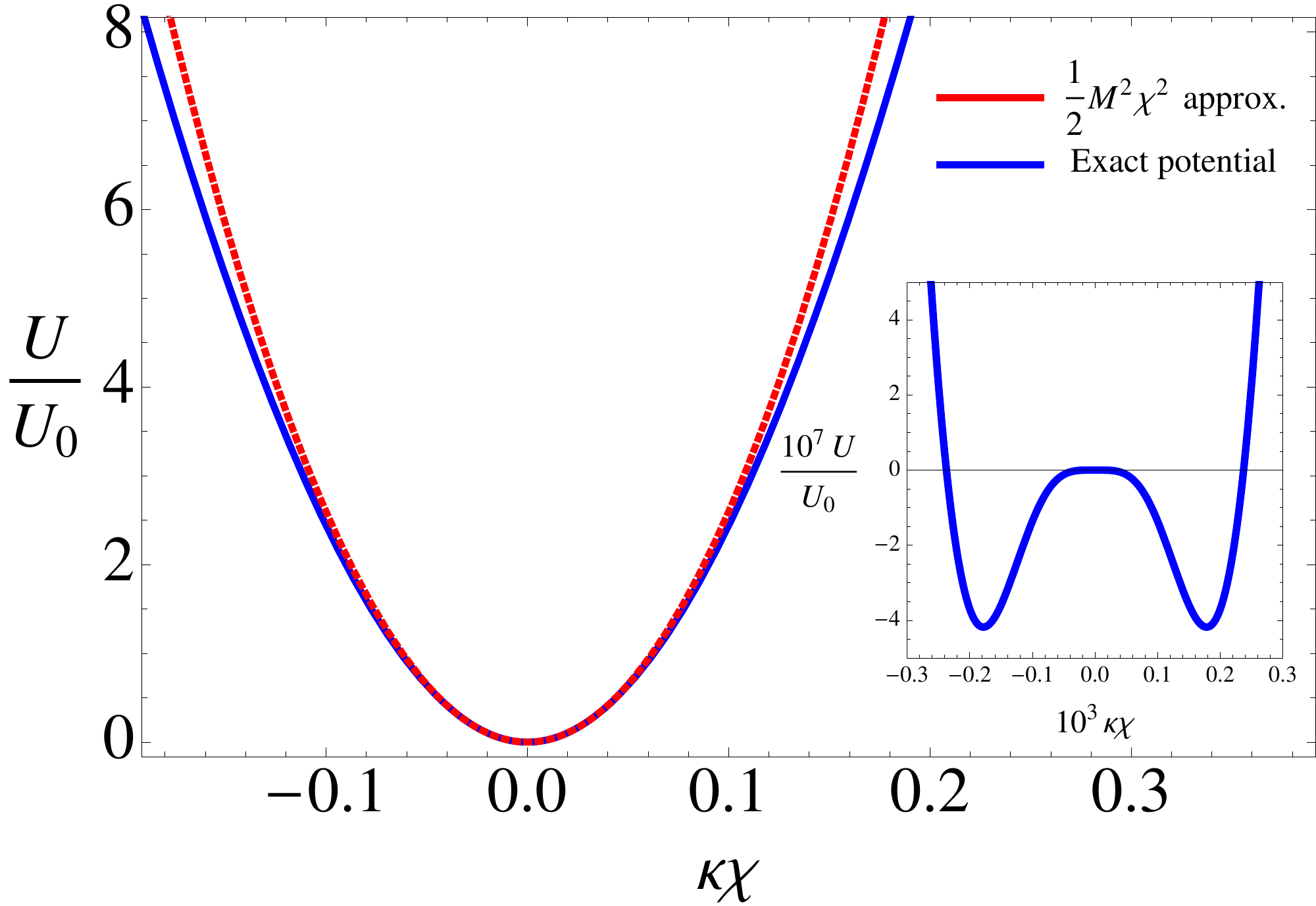}
\includegraphics[scale=0.43]{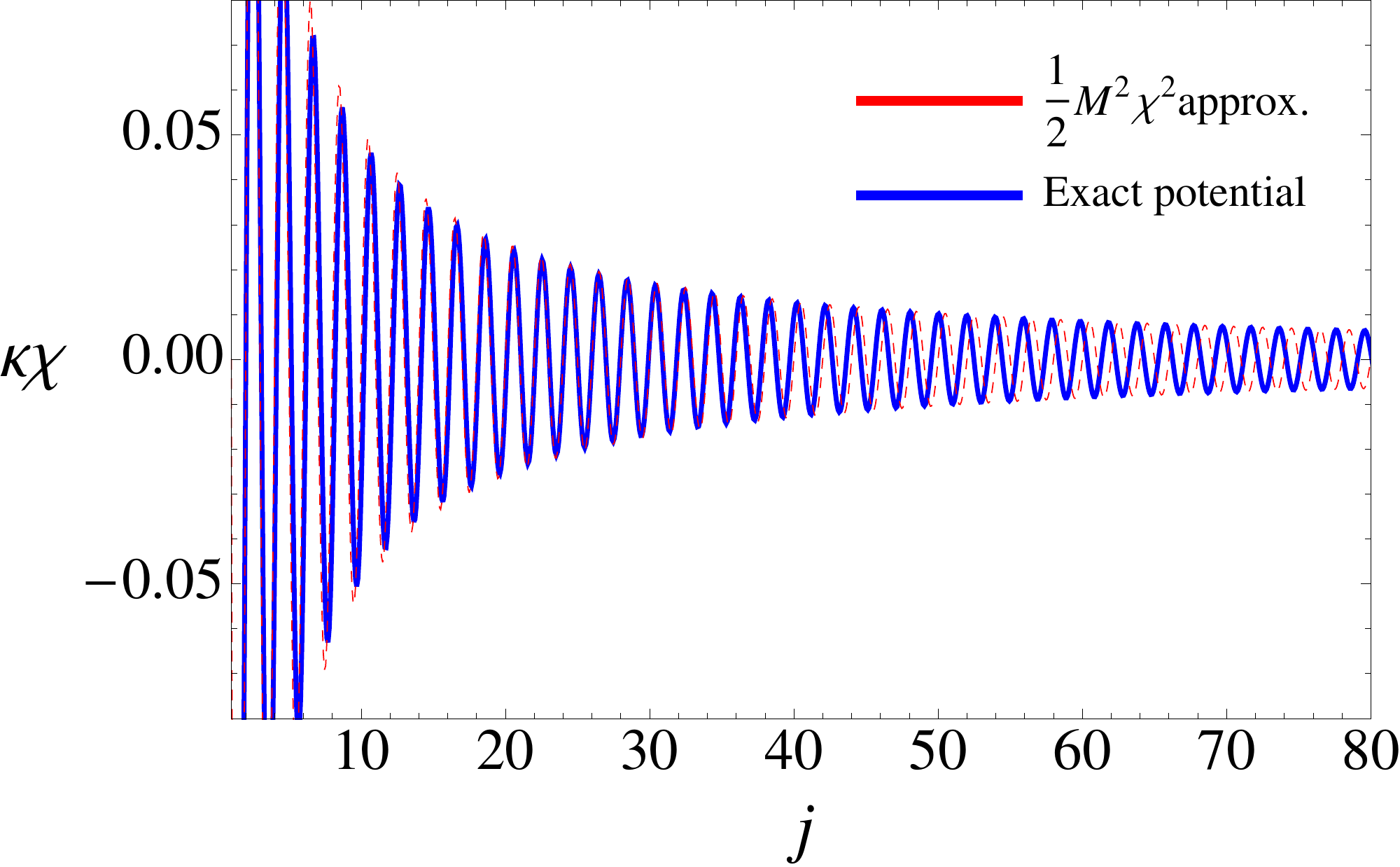}
\caption{(color online)  Top: Comparison between the exact renormalization group enhanced potential in the noncritical case and the quadratic approximation \eqref{potQ}. The normalization scale $U_0$ is taken to be $U_0=(10^{-3} M_P)^4$ and $\kappa=M_P^{-1}$..
Bottom: Evolution of the background field $\chi$ in the noncritical case as a function of the number of semioscillations $j$. Field values are measured in Planck units ($\kappa=M_P^{-1}$). }\label{potfig}
\end{figure}
 \subsection{Noncritical Higgs inflation}\label{sec:preh1}

 As shown in Fig.~\ref{potfig}, the RGE potential for noncritical Higgs inflation can be well approximated by a quadratic potential, except for the very small field values $\chi\sim {\cal O}(10^{-4}) M_P$ in which the self-coupling of the Higgs field becomes negative. For a large number of semioscillations, the evolution of the Higgs field is completely unaffected by the features of the potential at small field values. This allows us to apply the combined preheating formalism presented in Appendix \ref{appendix2}.
 
\begin{figure}
\centering
\includegraphics[scale=0.42]{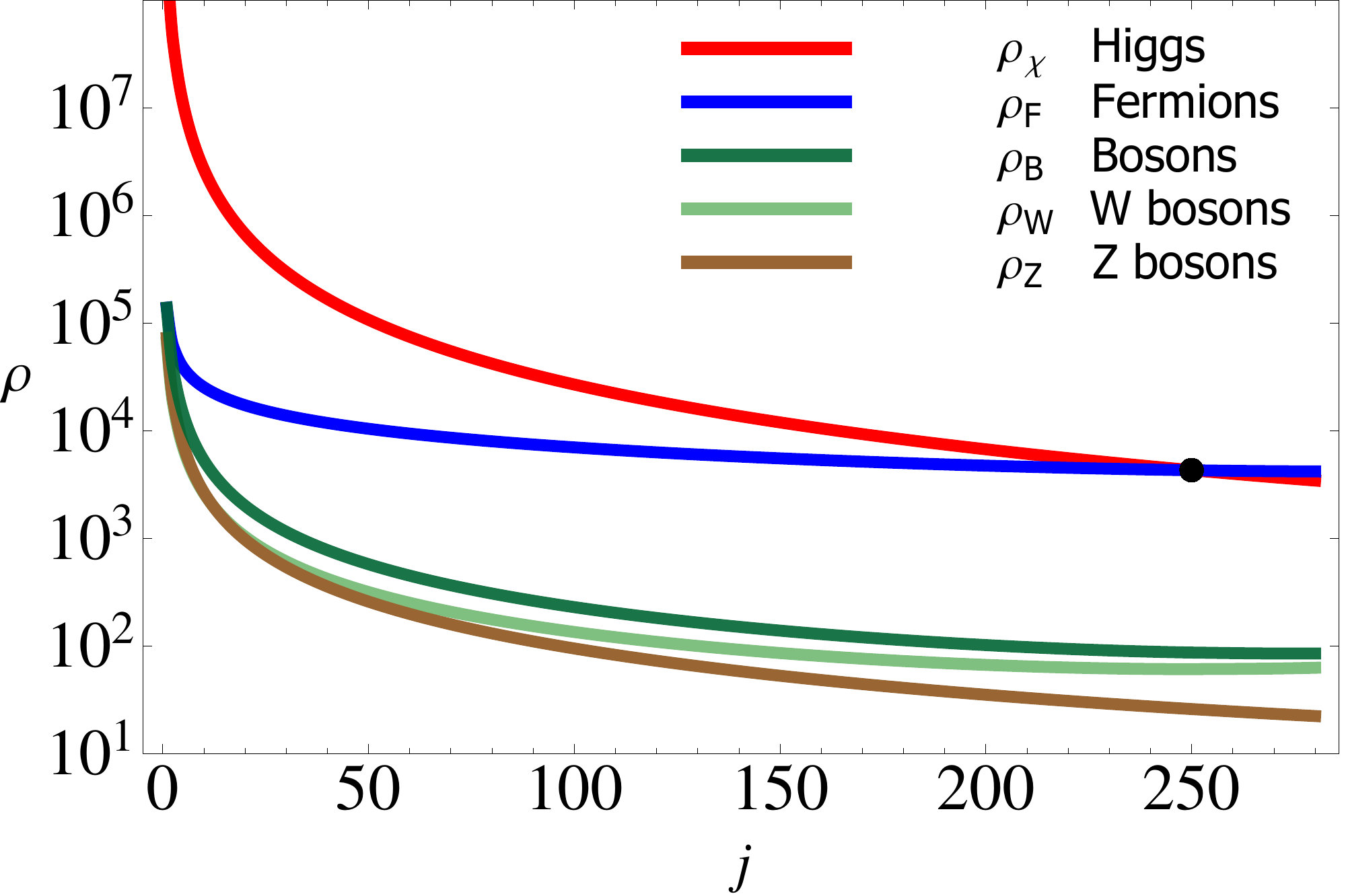}
\caption{(color online)  Evolution of the different energy densities (in $M^4$ units) for the noncritical case ($\lambda=3.4\times 10^{-3}$, $\xi=1500$, $g_1=0.44$, $g_2=0.53$) as a function of the number of semioscillations $j$.}\label{rhofig}
\end{figure}

 The energy densities for the created gauge bosons and fermions\footnote{We take $\lambda=3.4\times 10^{-3}$, $\xi=1500$, $g_1=0.44$,$g_2=0.53$.} [see~Eqs.~\eqref{rhoB}, \eqref{rhowpm}, \eqref{rhoz} and \eqref{rhoF}] as a function of the number of semioscillations $j$ are presented in Fig.~\ref{rhofig}. The production of $W$ and $Z$ particles in the first semioscillation is significantly larger than in the tree-level case\footnote{Indeed, for the first semioscillation and typical values of the couplings in the two cases, we have 
$\frac{\Delta n^\text{r}_F}{ \Delta n^\text{t}_F}\simeq\sqrt{\frac{\lambda_\text{r}}{\lambda_\text{t}}} \left(\frac{\xi_\text{t}}{\xi_\text{r}}\right)^{2}\simeq 20.5\,, \frac{E^\text{r}_F}{E^\text{t}_F}\simeq\sqrt{\frac{\xi_\text{t}}{\xi_\text{r}}}  \simeq 3.2$ and  
$\frac{\Delta \rho^\text{r}_F}{ \Delta \rho^\text{t}_F}\simeq\sqrt{\frac{\lambda_\text{r}}{\lambda_\text{t}}}\left(\frac{\xi_\text{t}}{\xi_\text{r}}\right)^{5/2}\simeq 65$ for the first semioscillation. Combining these results with the relation between the background energy densities in the two cases,
$\frac{\rho^\text{r}_\chi}{  \rho^\text{t}_\chi}\simeq\frac{\lambda_\text{r}}{\lambda_\text{t}}\left(\frac{\xi_\text{t}}{\xi_\text{r}}\right)^{2}\simeq 3.95\,,$
we get $\frac{\Delta \rho^\text{r}_F}{\rho^\text{r}_\chi}\simeq 15\frac{\Delta \rho^\text{t}_F}{ \rho^\text{t}_\chi}\,.$
}. The continuous production and subsequent decay rapidly sustains the energy density of the fermions against the expansion of the Universe.  Radiation domination takes place after $j_r=250$ semioscillations and precedes the onset of parametric resonance. The backreaction of the gauge bosons and fermions \textit{on the effective oscillation frequency} can be completely neglected at that time (see\ Appendix \ref{appendix3}).

\begin{figure}
\centering
\includegraphics[scale=0.42]{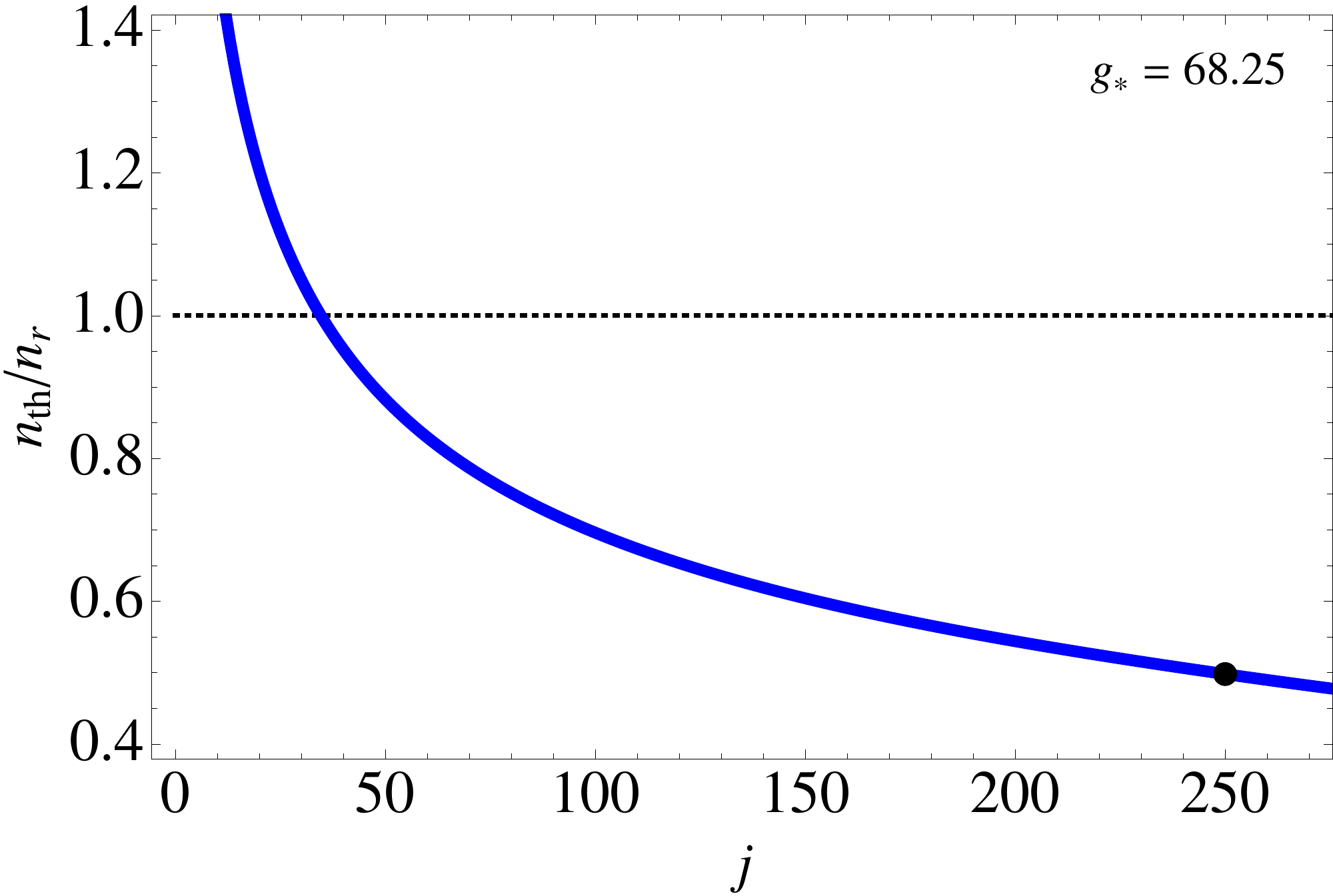}
\includegraphics[scale=0.42]{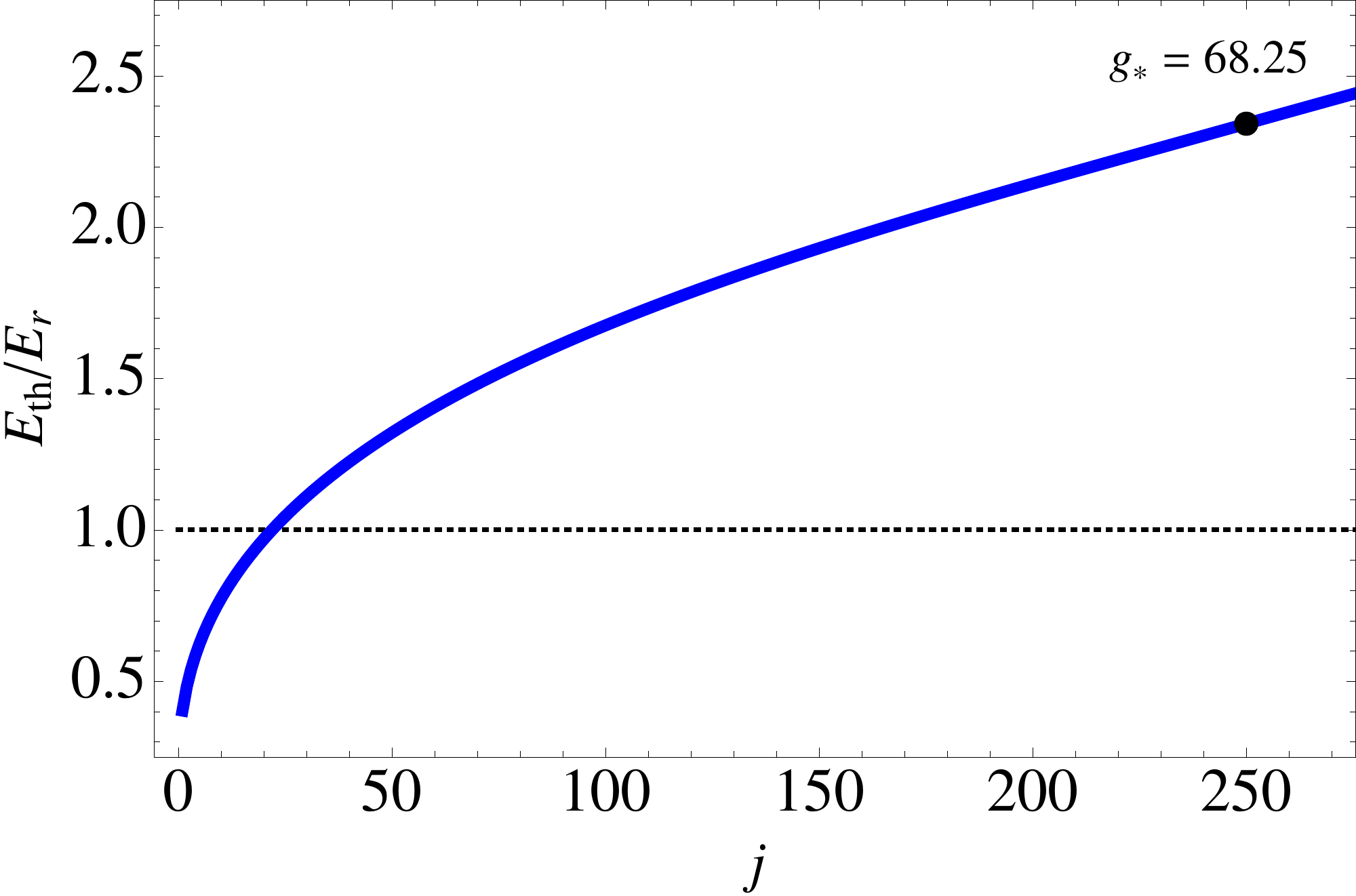}
\caption{Comparison between the number  and average energy of fermions and those associated to a thermal distribution with temperature $T_r(j)$ and total number of degrees of freedom $g_*=68.25$.}
\label{fig:nf}
\end{figure}

\begin{figure}
\centering
\includegraphics[scale=0.43]{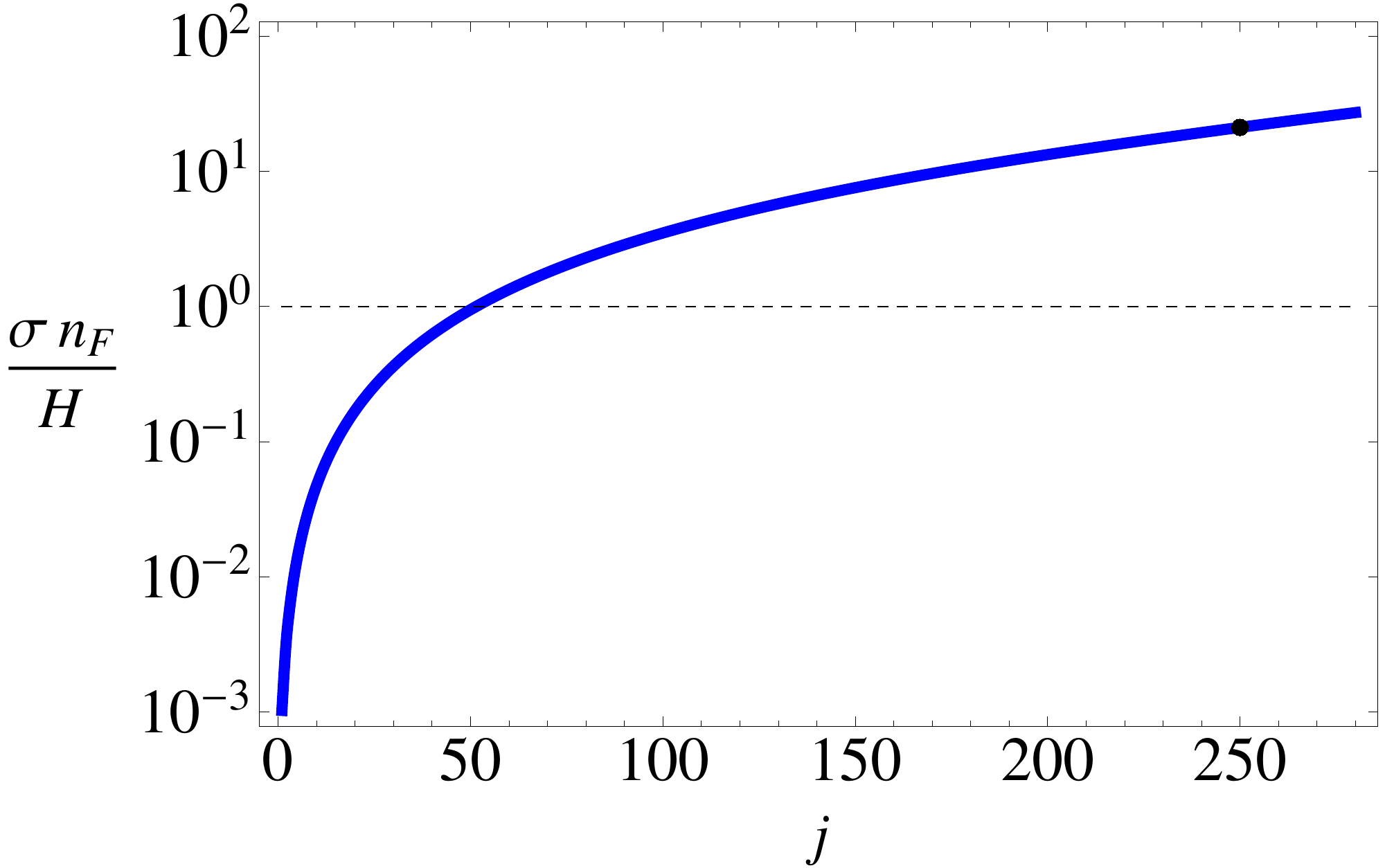}
\caption{Ratio between the fermion annihilation rate and the expansion rate as a function of the number of semioscillations $j$.}
\label{fig:cross}
\end{figure}

As shown in Fig.~\ref{fig:nf}, soon after the beginning of preheating, the sustained evolution of the total number of fermions due to particle creation together with the redshift of energies $E_{F(\text{B})}$ due to the expansion of the Universe drives the system into an overoccupied state made of low energetic particles with respect to those in a thermal distribution (see\ Eqs.~\eqref{nth} and \eqref{Eth}).

In the case of an overoccupied plasma, thermalization proceeds by energy cascading from the overoccupied modes with low momentum to the higher momentum modes with lower occupancy. Number-changing processes $2\leftrightarrow 1$ are expected to be parametrically as efficient as elastic scatterings or annihilations \cite{Kurkela:2011ti}. An estimate\footnote{A proper analysis would require the use of Boltzmann equations.} of the thermalization time can be obtained by comparing the annihilation rate of fermions via electroweak interactions\footnote{Here $\alpha_W\equiv g_2^2/(4\pi)$ is the electroweak coupling constant, $s\simeq\left(2E_F(j)\right)^2$ stands for the square of the center-of-mass energy and $n_F$ is the total number density of fermions.  We omit factors correcting for the charge of the fermion and the number of colors. The cross section for the annihilation through a gluon can be obtained (up to color factors), by simply replacing $\alpha_W$ by $\alpha_s$. The contribution of electroweak and QCD processes is expected to be quite similar due to the approximate unification of 
 $\alpha_W$ and $\alpha_s$ at the preheating scale.}
\begin{equation}\label{cross}
\Gamma\sim \sigma n_F \sim \frac{\alpha_W^2}{s} n_F\,,
\end{equation}
with the expansion rate.
As shown in Fig.~\ref{fig:cross}, the rate \eqref{cross} exceeds the Hubble rate just a few tens of oscillations after the end of inflation. The fermions present at radiation domination ($j=j_r$) will thermalize in $t_\text{th}\sim (\sigma n_F)^{-1}\vert_{j_r} \sim 0.05\,H^{-1}\vert_{j_r} \sim 15$ semioscillations.
This allows us to interpret the radiation temperature $T_r$ at that time as the reheating temperature $T_{RH}$ setting the onset of the hot big bang. Taking into account the number of degrees of freedom after thermalization ($g_*=106.75$), we obtain
\be
T^{NC}_{RH}\simeq 1.8 \times 10^{14}\hspace{2mm}\text{GeV}\,.
\ee 
This value exceeds the critical temperature $T^{NC}_\text{+}$ needed for driving the Higgs field towards the true electroweak vacuum [see\ Eq.~\eqref{TcritNC}]. Noncritical Higgs inflation can take place even if the SM vacuum is metastable.

\subsection{Critical Higgs inflation}\label{sec:preh2}
\begin{figure}
\includegraphics[scale=0.4]{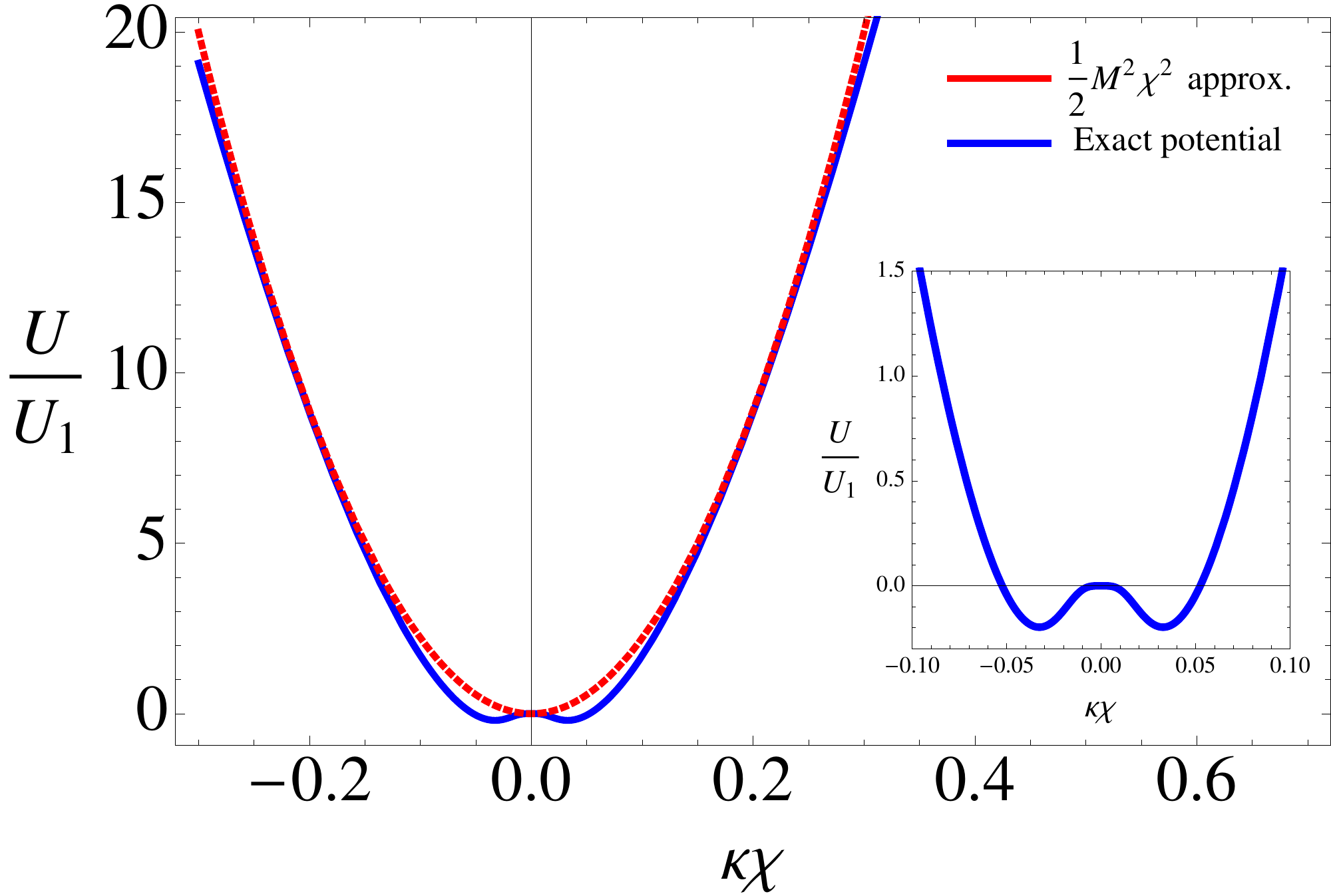}
\includegraphics[scale=0.41]{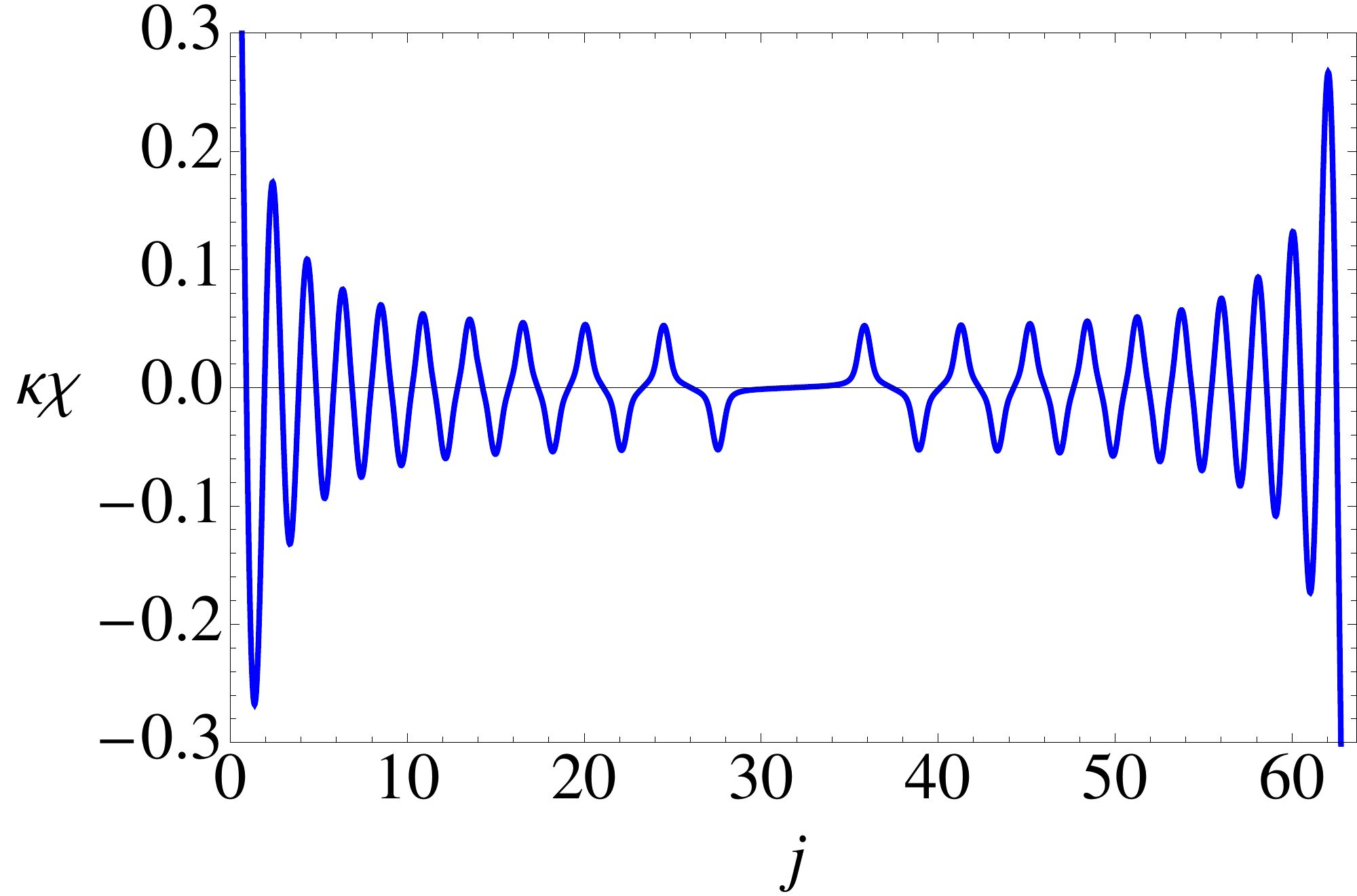}
\caption{(color online) Top: Comparison between the exact renormalization group enhanced potential and the quadratic approximation \eqref{potQ} in the critical case. The normalization scale $U_1$ is taken to be $U_1=10^{-9} M_P^4$ and $\kappa=M_P^{-1}$. Bottom: Evolution of the background field $\chi$ in the critical case as a function of the number of semioscillations $j$.}
\label{fig:potcrit}
\end{figure}
In critical Higgs inflation, the value of the non-minimal coupling $\xi$ is relatively small ($\xi \sim 10$) and the jumps in the coupling constants appear closer to the Planck scale (see\ Fig.~\ref{lmu}). As shown in Fig.~\ref{fig:potcrit}, the energy stored in the Higgs field after inflation is {\em comparable} to the height of the barrier separating the two vacua. Converting the energy density of the inflaton field at the end of inflation ($V^{1/4}\simeq 6\times 10^{16}$~GeV) into an instantaneous radiation temperature, we obtain an upper bound $T^C_\text{max}$ on the reheating temperature\footnote{A more realistic bound taking into account the particle production at the bottom of the potential is presented, for completion, in Appendix \ref{appendix4}.}
\begin{equation}
T^C_\textit{RH}<T^C_\text{max}= 7.85 g_*^{-1/4} \times 10^{16}\,\, \text{GeV}\,.
\end{equation}

\begin{figure}[h!]
\includegraphics[scale=0.4]{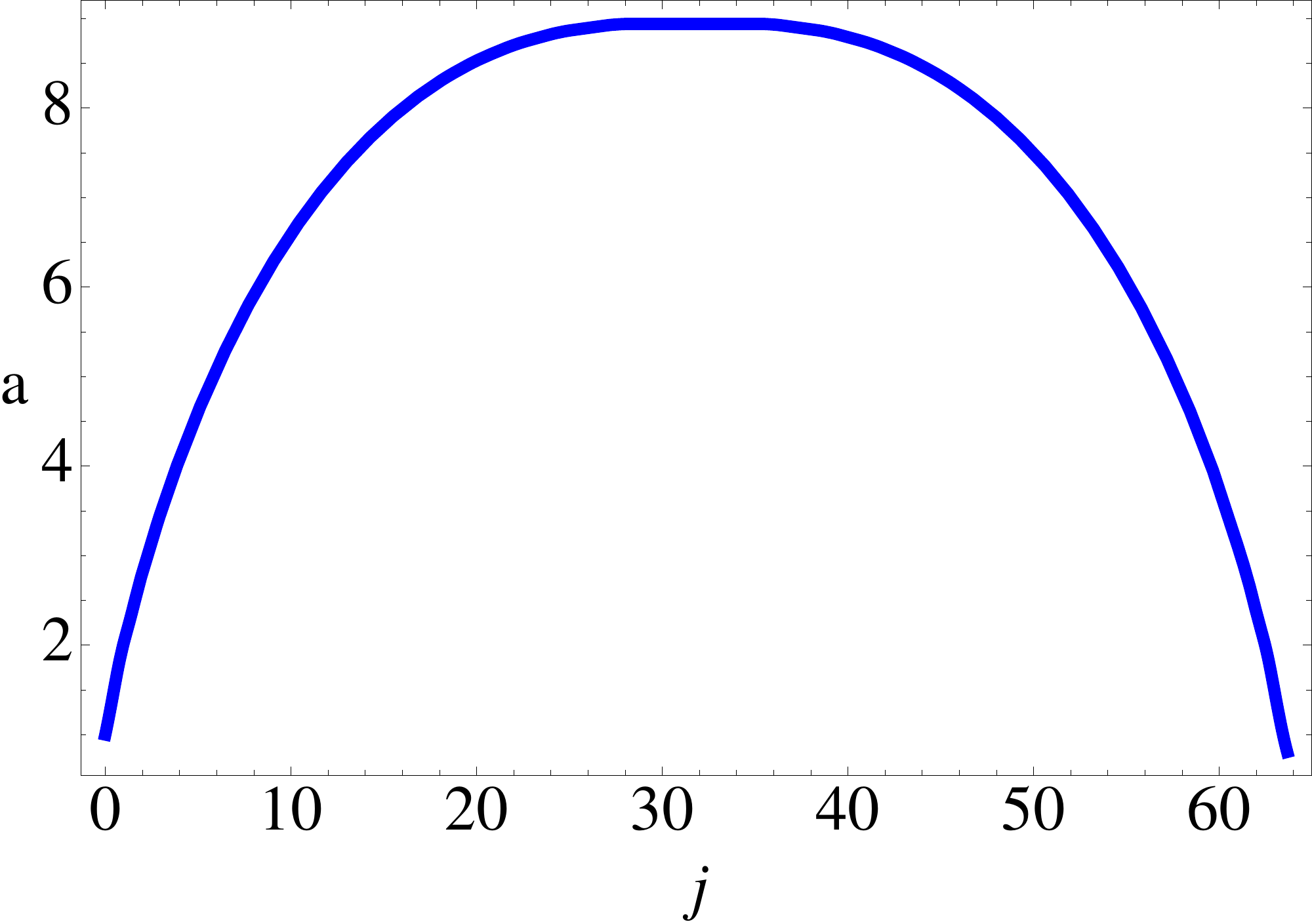}
\caption{Scale factor as a function of the number of semioscillations $j$.}
\label{fig:acrit}
\end{figure}

Since $T^C_\textit{RH}<T^C_\text{max}< T_+$, the shape of the potential remains unchanged in the presence of thermal corrections and the system inevitably relaxes towards the minimum of the potential at Planck values. When the energy on the field becomes equal to the amplitude of the barrier ($j\simeq 30$), the expansion of the Universe stops. From there on, the scale factor begins to shrink and the amplitude of the field increases with time (see\ Fig.~\ref{fig:potcrit}). Eventually, the kinetic energy of the Higgs field starts dominating the total energy density $(p\approx \rho)$ and the Universe generated by critical Higgs inflation with metastable electroweak vacuum collapses \cite{Felder:2002jk}, see\ Fig.~\ref{fig:acrit} . 

\section{Conclusions}\label{sec:con}

Although the present experimental data are perfectly consistent with the absolute stability of Standard Model within the experimental and theoretical uncertainties, one should not exclude the possibility that other experiments will be able to establish the metastability of the electroweak vacuum in the future. Should the Higgs inflation idea be abandoned in this case? This paper gives a negative answer to this question.

We reconsidered the validity of Higgs inflation for values of the Higgs and top quark masses giving rise to the instability of the SM vacuum at energy scales below the scale of inflation.  The nonminimal coupling to gravity makes the SM nonrenormalizable and requires the addition of an infinite number of counterterms. There are several ways to do this. Let us consider for concreteness the Einstein frame. The most general approach would be to include all sorts of higher-dimensional operators suppressed by the Planck scale. However this will automatically destroy the whole class of large field inflationary models to which Higgs inflation belongs\footnote{Note however that this is not so in Higgs inflation if these operators are added in the Jordan frame.}. To avoid these well-known problems, we adopted a minimalistic approach in which the structure of the counterterms is restricted by several self-consistent hypotheses concerning the symmetries of the UV completion, which is assumed to be scale invariant. The added counterterms differ from those already present  in the original theory but do not  modify the asymptotic properties of the model. In particular, the evolution equations for the coupling constants can be approximated by the usual SM renormalization group equations at low energies and by those of the chiral Standard Model at high energies. The ambiguities associated with the nonrenormalizability of the theory appear only in the narrow interface between these two asymptotic regions and are connected to the finite parts of the counterterms. These finite parts give rise to jumps in the evolution of the coupling constants, the amplitude of which cannot be determined within the theory itself. We will like to emphasize that the idea presented on this paper \textit{does not} depend on the particular order at which the set of RG equations is truncated. Within the minimalistic approach to Higgs inflation, all threshold effects (independently of their order) happen at the same energy scale. The validity of the scenario depends only on the \textit{collective effect} of all these thresholds, and in particular on their ability to convert a negative Higgs self-coupling below the transition scale into a positive one at the inflationary scale.

{The general ideas presented in this paper extend beyond the Higgs inflation scenario and could be applied to other models of inflation not directly driven by the Higgs field. Surviving the instability of the EW vacuum requires the following ingredients i) whatever the new physics beyond the Standard Model, it should be able to stabilize the Higgs potential below the scale of inflation  ii)  the resulting potential should lay on top of the Standard Model one iii) whatever the mechanism giving rise to inflation, the reheating process should be efficient enough to backreact into the Higgs potential and restore the symmetry.
As a proof of existence, we determined a set of parameters giving rise to noncritical Higgs inflation in the case of a metastable SM vacuum and studied the subsequent evolution of the Universe. Taking into account the nonperturbative production of SM particles after the end of inflation, we estimated the reheating temperature and compared it with the temperature needed to stabilize the effective potential. We showed that, while critical Higgs inflation does necessarily require the absolute stability of the SM vacuum, the successful noncritical Higgs inflation can be possible even if our vacuum is metastable. 

\begin{acknowledgements}
This work was partially supported  by the Swiss National Science
Foundation. We  thank  Igor Tkachev for useful discussions.
\end{acknowledgements}

\begin{appendix}

\section{Truncation of the RG equation chain} \label{sec:two loop}

This appendix closely follows the general arguments of Ref.~\cite{Bezrukov:2010jz}, particularizing them to the present setup.

The counterterms \eqref{counterL} and \eqref{counteryt} have a functional dependence on the background field $\chi$ that differ from that in the tree-level Lagrangian. 
The consistency of the computation requires the inclusion of these terms in the original Lagrangian and the reevaluation of the radiative corrections. The one-loop diagrams associated to the new pieces are given by\footnote{We retain just the first term in Eq.~\eqref{counteryt} because of its nontrivial contribution at small $\chi$. The second term contributes only for intermediate intermediate field values. 
}
\begin{eqnarray}
  \tikz[baseline=-0.5ex]{\hspace{-1cm}
    \draw[higgsline] (0,0em) circle (1.4em);
    \draw[fill] (0,-1.4em) circle (0.3em)
    node[below] {$\delta\lambda[(F'^2+\frac{1}{3}F''F)^2F^4]''$};\label{A1}
  }
  &\hspace{-1.9cm}\sim   \,\,& \delta\lambda[(F'^2+\frac{1}{3}F''F)^2F^4]'' \lambda(F^4)''
  \\
  \tikz[baseline=-0.5ex]{\hspace{-2.1cm}\label{A2}
    \draw[fermionline] (0,-1.4em) arc (270:-90:1.4em);
    \draw[fill] (0,-1.4em) circle (0.3em)
    node[below] {$\delta yF'^2F$};
  }
  &\hspace{-1.9cm}\sim  \,\,&\delta yF'^2F y^3F^3. 
\end{eqnarray}
These contributions should be compared with the two-loop contributions generated by the original Lagrangian\footnote{The particular combination in Eq.~\eqref{A5} is chosen for illustrative purposes. Provided there are two masses present in the propagators, other combinations involving $(yF)^2$ and $\lambda(F^4)''$ can also appear.
}
\begin{eqnarray}
 \label{eq:2loop-diag}
  \tikz[baseline=-0.5ex]{\hspace{-0.5cm}
    \draw[higgsline] (0,0) circle (1.4em);
    \draw[higgsline] (-1.4em,0) -- (1.4em,0);
    \node[left] at (-1.5em,0) {$\lambda(F^4)'''$};
    \node[right] at (1.5em,0) {$\lambda(F^4)'''$};
  }
 &\hspace{-0.5cm} \sim \,\,& (\lambda(F^4)''')^2 \lambda(F^4)''\,,\label{A3}
  \\
   \tikz[baseline=-0.5ex]{\hspace{-0.5cm}
    \draw[higgsline] (0,1.4em) circle (1.4em);
    \draw[higgsline] (0,-1.4em) circle (1.4em);
    \node[right] at (0.5,0) {$\lambda(F^4)^{''''}$};
  }
 &\hspace{-0.5cm} \sim  \,\,&\lambda(F^4)^{''''} (\lambda(F^4)'')^2\,,
  \\
   \tikz[baseline=-0.5ex]{\hspace{-1.1cm}
    \draw[fermionline] (0,-1.4em) arc (270:-90:1.4em);
    \draw[higgsline] (-1.4em,0) -- (1.4em,0);
    \node[left] at (-1.5em,0) {$y F'$};
    \node[right] at (1.5em,0) {$y F'$};
  }
  &\hspace{-0.5cm}\sim  \,\,& (y F')^2 (y F)^4\,,\label{A5}
  \\
  \tikz[baseline=-0.5ex]{\hspace{-1.1cm}
    \draw[fermionline] (0,0) arc (270:-90:1.4em);
    \draw[higgsline] (0,-1.4em) circle (1.4em);
    \node[right] at (0.5,0) {$y F''$};
  }
  &\hspace{-0.5cm}\sim  \,\,& y F'' (yF)^3 \lambda (F^4)''\,,\label{A6}
\end{eqnarray}
A simple inspection of Eqs.~\eqref{A1}-\eqref{A6} reveals that in order to have a good perturbative expansion the finite parts $\delta\lambda$ and $\delta y_t$ must be of the same order (in power counting) than the loop corrections producing them (see~Sec.~\ref{subset:C})
\begin{equation}
  \label{eq:hier1}
  \delta \lambda \sim O(\lambda^2,y^4),
  \quad
  \delta y \sim O(y^3,y \lambda)\,.
\end{equation}
This condition, together with the standard power-counting assumption $\lambda\sim {\cal O}(y^2)$, ensures that the \textit{one-loop} contributions generated by the counterterms 
\eqref{A1} and \eqref{A2} are of the same order than the \textit{two-loop} contributions \eqref{A3}-\eqref{A6} generated by the tree-level Lagrangian.

The previous recipe can be easily generalized to higher orders. For instance, since the contributions of the diagrams \eqref{A1}-\eqref{A6} are of order
$\lambda^3, \lambda^2y^2,\lambda y^4,y^6$, these would be the orders that should be assumed for the finite parts of the new counterterms to be added in order to cancel the associated divergencies.
If the finite parts of the counterterms introduced in each iteration are hierarchical, the loop
expansion can be consistently truncated at a given loop order.

\section{Combined preheating formalism} \label{sec:combined}
In this appendix, we summarize the combined preheating formalism \cite{  
  GarciaBellido:2008ab,Bezrukov:2008ut}. Let assume the shape of the inflationary potential during the whole reheating can be well approximated by a quadratic potential\footnote{As for instance happens in the noncritical case, see Sec.~\ref{sec:preh1}.}
\begin{equation}\label{potQ}
U(\chi) \simeq {1\over2}M^2\chi^2\,, \hspace{5mm} M = \sqrt{\frac{\lambda}{3}}\frac{M_P}{\xi}\,.
\end{equation}
In this potential, the Universe expands as in a matter-dominated background ($a\propto t^{2/3}$) with zero pressure and energy 
density $\rho_\chi(t)=\frac{1}{2} M^2 \chi(t)^2$. The evolution of the Higgs field is given by
\begin{equation}\label{phievol}
\chi(t)=\frac{\chi_e\sin(M t)}{M t }=\frac{\chi_e\sin(\pi j)}{\pi j}\equiv \chi(j)\sin(\pi j)\,,
\end{equation}
with $j=Mt/\pi$ the number of semioscillations or zero crossings and  $\chi_e=\sqrt{8/3}M_P$ an initial amplitude dictated by the covariant conservation law $\dot \rho_\chi=-3 H \rho_\chi$.

The evolution equation for the gauge boson fluctuations in this background is 
\begin{equation}
\ddot B_k+3 H\dot B_k+\left(\frac{k^2}{a^2}+ \tilde m_B^2(t)\right) B_k =0\;, 
\end{equation}
with $B=W,Z$ and 
\begin{equation}\label{masses}
\tilde{m}^2_B (t)  \equiv \frac{{m_B}^2}{\Omega^2}=\frac{g^2 M_P^2(1-e^{-\sqrt{2/3}\vert\chi(t)\vert/M_P})}{4\xi}\,,
\end{equation}
the conformally rescaled version of Jordan frame masses $m_B=gh/2$. The friction term $3 H\dot B_k$ can be eliminated by performing a conformal redefinition of the gauge fields ($B_k\rightarrow a^{-3/2} B_k$) to obtain\footnote{The redefinition introduces terms proportional to $H^2$ and $\ddot a/a$ that can be safely neglected at scales below the horizon.}
\begin{equation}\label{evolM}
B_k'' +\left(K^2+ \frac{\tilde  m_B^2(t)}{M^2}\right) B_k =0\;, 
\end{equation}
with $K \equiv \frac{k}{aM}$ a rescaled momentum and the primes denoting derivatives with respect to a rescaled time $\tau = Mt$. Here we have adopted a compact notation in which $g=g_2,g_2/\cos\theta_W$ for the $B=W,Z$ bosons respectively, $\theta_W=\tan^{-1}(g_1/g_2)$ is the weak mixing angle and $g_1$ and $g_2$ are the gauge couplings associated to the Standard Model  $U(1)_Y$ and $SU(2)_L$ gauge groups. 

Particle production takes place within a very restricted field interval $(\vert \chi\vert \ll\chi_a)$ around the minimum of the potential, in which the adiabaticity condition $\vert \dot{\tilde m}_B\vert\ll \tilde m_B^2$ is violated \cite{Kofman:1997yn}.  In this region, the effective square masses \eqref{masses} become linear in the field [$\tilde{m}^2_B\propto \vert\chi(\tau)\vert \propto\vert \sin \tau\vert/j \approx \vert \tau\vert/j$] and the evolution equation \eqref{evolM} can be rewritten as
\begin{equation}\label{evolB}
-B_k'' - \frac{q_B}{j}|\tau| B_k = K^2 B_k\;, \hspace{10mm} q_B \equiv \frac{g^2\xi\,}{\pi\lambda}\;.
\end{equation}
The previous equation can be formally interpreted as the Schr\"odinger equation of a particle crossing a (periodic) inverted triangular potential. To solve it, we note that the nonadiabaticity region $\chi_a= \left[\lambda\pi j/(4 g^2\xi)\right]^{1\over3}\,\chi(j)\simeq (10^{-6} j)^{1/3}\,\chi(j)$ is much smaller than the background field value $\chi(j)$ for at least hundred thousand oscillations. This justifies the use of a WKB approximation for computing the number of particles after the $j$-th scattering, $n_{k}(j^+) $, in terms of the number of particles just before that scattering, $n_{k}(j^-)$. After some computations, we get \cite{GalindoPascualbook}
\begin{eqnarray}\label{nk}
n_{k}(j^+) &= & C(x_j) + \left(1+2C(x_j) \right)n_{k}(j^-) \\ &+&2\cos\theta_{j-1} {\sqrt{C(x_j)\left[C(x_j)+1\right]}}\sqrt{n^2_{k}(j^-)+n_{k}(j^-)}\,, \nonumber
\end{eqnarray}
with 
\begin{equation}
C(x_j) \equiv  \pi^2\left[{\rm{Ai}}\left(-x_j^2\right){\rm{Ai}}'\left(-x_j^2\right) + {\rm{Bi}}\left(-x_j^2\right){\rm{Bi}}'\left(-x_j^2\right)\right]^2\,, \nonumber
\end{equation}
an infrared window function depending on Airy functions of first and second type,
\begin{equation}
x_{j} \equiv \frac{K}{(q_B/j)^{1/3}}=\frac{j^{1/3}k}{Mq_B^{1/3}a_j}\,,
\end{equation}
and  $\{\theta_j\}$ some accumulated phases at each scattering. A simple estimation of these phases reveals that they are essentially incoherent  [$\Delta \theta_j \sim g\left(\xi/\lambda\right)^{1/2}j^{-1/2}\sim {\cal O}(10^2)j^{-1/2} \gg \pi$] for the first few thousands of oscillations (see\ Ref.~\cite{GarciaBellido:2008ab,Bezrukov:2008ut} for details). This allows us to reduce \eqref{nk} to a phase-average relation \cite{Figueroa:2009jw}
\begin{equation}\label{nk2}
\left(\frac{1}{2}+n_{k}(j^+)\right) \simeq A(x_j)\left(\frac{1}{2}+n_{k}(j^-)\right)\,, 
\end{equation}
with enhancing Bose factor $A(x_j)\equiv1+2\text{C}(x_j)$.

Once produced, the gauge bosons tend to transfer energy into the Standard Model fermions ($F$) through decays ($B\rightarrow F\bar F$) and  annihilations ($B B\rightarrow F\bar F $). The decay modes are expected to be the dominant processes at early times, where the number densities are still low ($ \Gamma_B \gg \sigma n_B$). Let us assume that this relation holds for the typical number of semioscillations we are interested in\footnote{As we will show \textit{a posteriori} in Appendix \ref{appendix2}, this in indeed a very good approximation.}. In that case, the occupation numbers for the gauge bosons just before the $j$-th scattering can be written as
\begin{equation}\label{nk3}
n_k(j^-) =n_k((j-1)^+)e^{-\langle\Gamma_B\rangle_{j-1}\frac{T}{2}}\;.
\end{equation} 
The average $\left\langle\Gamma_B\right\rangle_j$ stands for the mean decay width of the $W$ and $Z$ bosons between two consecutive zero-crossings\footnote{Note that the number of $W$ bosons surviving every semioscillation of the Higgs field is larger than the number of $Z$ bosons ($\Gamma_W < \Gamma_Z$). The $W$ bosons are expected to become resonant before the $Z$ bosons do.} \cite{Cheng:2006book}
\begin{eqnarray}\label{decayWZ}
\left\langle\Gamma_{W\rightarrow all}\right\rangle_j &=& \frac{3g_2^2\langle \tilde{m}_W\rangle}{16\pi}\equiv \frac{2\gamma_W}{T} F(j)\;,\\
\left\langle\Gamma_{Z\rightarrow all}\right\rangle_j &=& \frac{2\text{Lips}\left\langle\Gamma_{W\rightarrow all}\right\rangle_j }{3\cos^3\theta_W}\equiv \frac{2\gamma_Z}{T} F(j)\,,
\end{eqnarray}
with $\text{Lips}\equiv\frac{7}{4}-\frac{11}{3}\sin^2\theta_W+\frac{49}{9}\sin^4\theta_W$ a \textit{Lorentz-invariant phase-space} factor,
\begin{equation}
\gamma_W \equiv\frac{3g_2^{3}}{32}\left(\frac{3\xi}{\lambda}\right)^{1/2} \,,\hspace{3mm}
 \gamma_Z \equiv \frac{2\text{Lips}}{3\cos^3\theta_W}\gamma_W\,.
\end{equation}
and\footnote{The last equality in this equation is just a good fit for \textit{all} $j$, including the first semioscillations.}
\begin{equation}
\label{Fj}
F(j) \equiv \int_{0}^{\pi} \frac{dx_j}{\pi}\left(1-e^{-\sqrt{2/3}|\chi(x_j)|/M_P}\right)^{1/2} \nonumber \simeq \frac{1}{0.57 + 1.94\sqrt{j}}\,.
\end{equation}
\noindent Combining Eqs.~\eqref{nk2} and \eqref{nk3}, we obtain \cite{Figueroa:2009jw}
\begin{eqnarray}\label{iternk}
\hspace{-11mm}\left(\frac{1}{2}+n_k((j+1)^+)\right)\hspace{-1mm}=\hspace{-1mm}A(x_j)\left(\frac{1}{2}+n_k(j^+)\,e^{-\gamma F(j)}\right)
\end{eqnarray}
The recursive iteration of this master equation allows us to obtain the total number of  gauge bosons at each crossing 
\begin{eqnarray}\label{nB}
\hspace{-12mm}n_B(j^+)=\hspace{-1mm}\int\frac{\,k^2 n_k(j^+)dk}{2\pi^2 a_j^3} = \frac{q_B M^3}{2\pi^2 j} \hspace{-1mm} \int \hspace{-1mm}  x_j^2 \,n_k(x_j^+) dx_j\,,
\end{eqnarray}
and with it their total energy density\footnote{The factor 2 accounts for the $W^+$ and $W^-$ while the factor 3 reflects the fact that each gauge boson can have one of three polarizations.}
\begin{eqnarray}
\rho_B(j) &=&\rho_W(j) + \rho_Z(j)\label{rhoB}\,,\\
\rho_W(j) &=& 2\times 3\, n_W(j^+)\langle m_W \rangle_{j}\,, \label{rhowpm} \\
\rho_Z(j) &=&1\times 3\,n_Z(j^+)\langle m_Z \rangle_{j} \label{rhoz}\,.
\end{eqnarray}  
On the other hand, the number of fermions produced by the decay of the gauge bosons between two consecutive scatterings and their corresponding energy density are given by
  \begin{eqnarray}
\hspace{-10mm}&&  \Delta n_{F}(j)\equiv \Delta n_{F}^{(\rm W)}(j)+\Delta n_{F}^{(\rm W)}(j)\,, \\
\hspace{-10mm}&&\Delta\rho_F{(j)} = \Delta n_{F}^{(\rm W)}(j) E_{F}^{(\text{W})}(j) +\Delta n_{F}^{(\rm Z)}(j) E_{F}^{(\text{Z})}(j) \,,
\end{eqnarray}
with\footnote{The factor 2 accounts for the fact that each gauge boson decays into two fermions.}
  \begin{eqnarray}
\hspace{-10mm}&&\Delta n_{F}^{(\rm W)}(j) =2\times 3   \left(2\,n_W(j^+)(1-e^{-\gamma_WF(j)})\right)\,, \label{nFw}\\ 
\hspace{-10mm}&&\Delta n_{F}^{(\rm Z)}(j) = 2\times 3   \left( n_Z(j^+)(1-e^{-\gamma_ZF(j)}) \right)\label{nFz}\,,
\end{eqnarray} 
and
\begin{equation}\label{Efermion}
E_{F(\text{B})}{(j)}\approx \frac{1}{2}\langle \tilde m_B \rangle_j=\frac{\sqrt{3 \pi q_B}}{4}  F(j)M \,,
\end{equation}
 the mean energy of the relativistic decay products $F$ of the nonrelativistic  gauge boson $B$. Summing over the number of semioscillations and taking into account the dilution due to the expansion of the Universe, the total number of fermions and their total energy density after $j$ semioscillations become
 \begin{equation}\label{ntF}
n_F{(j)} = \sum_{i=1}^{j} \left(\frac{i}{j}\right)^{2}\Delta n_{F}(i)\,,
\end{equation}  
\begin{equation}\label{rhoF}
\rho_F{(j)} = \sum_{i=1}^{j} 
\left(\frac{i}{j}\right)^{8/3}\Delta\rho_F{(i)}\,,
\end{equation}  

\section{Consistency checks} \label{appendix2}

The combined preheating formalism presented in the previous section assumes the following:
\begin{enumerate}[i.]
\item The frequency $M$ used to describe the background evolution in \eqref{evolM} is not significantly modified by particle production.
\item  The condition $ \Gamma_B \gg \sigma n_B$ holds during the whole preheating process.
\end{enumerate}
 To verify the consistency of the approach, we perform two consistency checks: 
\begin{enumerate}[i.]
\item
We use the number densities obtained through the combined preheating formalism to estimate the backreaction on $M$. The effective frequency is given by $\omega^2\approx M^2\left[1+\Delta M^{B}_{BR}+\Delta M^{F}_{BR}\right]$, with 
\begin{eqnarray}\label{br}
\Delta M^{B}_{BR} &\equiv& \frac{g\sqrt{\lambda}n_B(j) }{6^{3/4}}\left(\frac{\pi j }{\xi}\right)^{3/2} \,,\\ 
\Delta M^{F}_{BR} &\equiv& \frac{y_f\sqrt{\lambda}n_B(j) }{\sqrt{2}\,6^{3/4}}\left(\frac{\pi j }{\xi}\right)^{3/2} 
 \end{eqnarray}
the contribution of the created gauge bosons and fermions \cite{GarciaBellido:2008ab}. 

\begin{figure}
\centering
\includegraphics[scale=0.4]{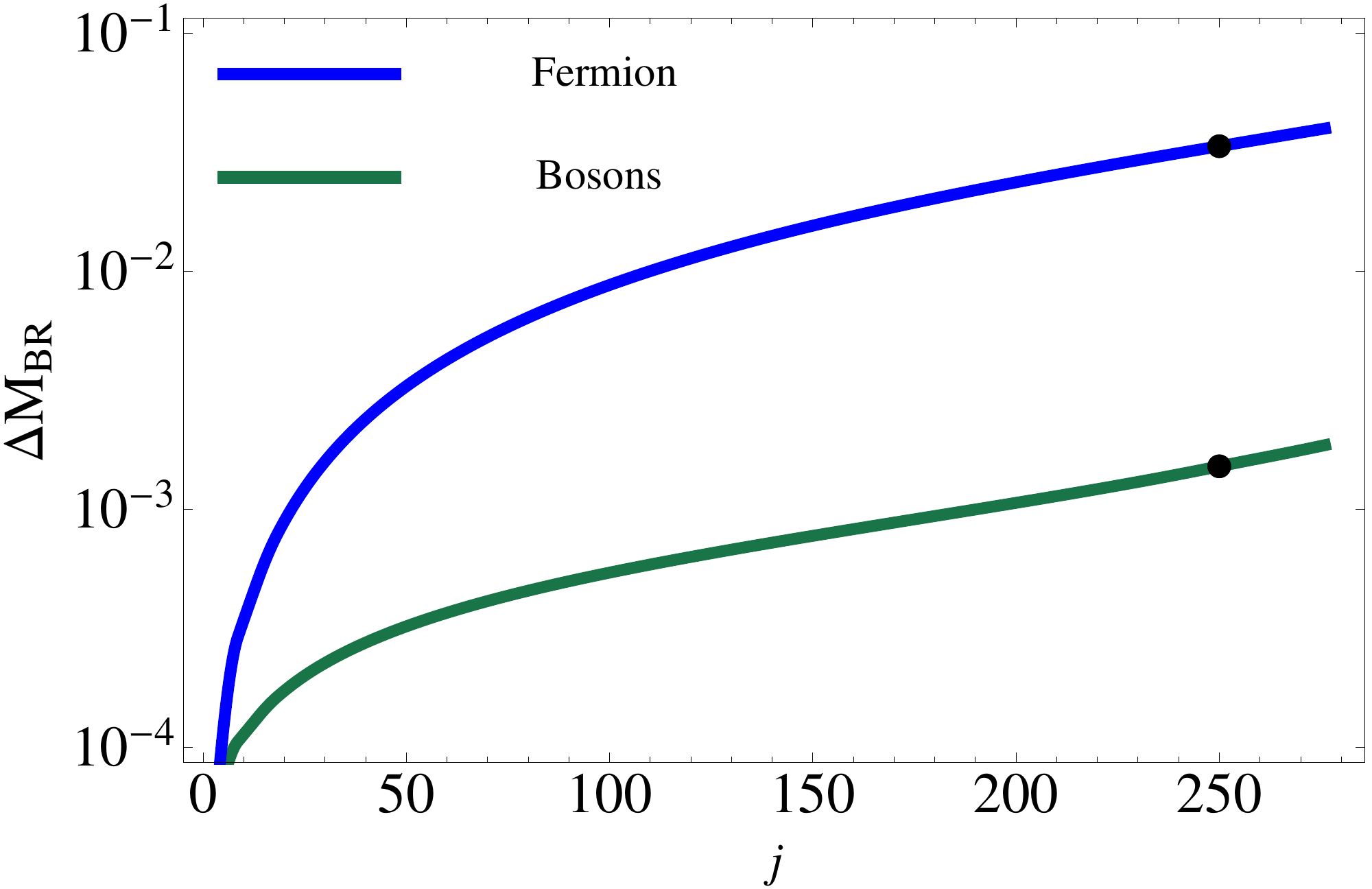}
\caption{(color online) Contribution of the created gauge bosons and fermions to the background frequency $M$ as a function of the number of semioscillations $j$.}
\label{brfig}
\end{figure}

\begin{figure}
\centering
\includegraphics[scale=0.40]{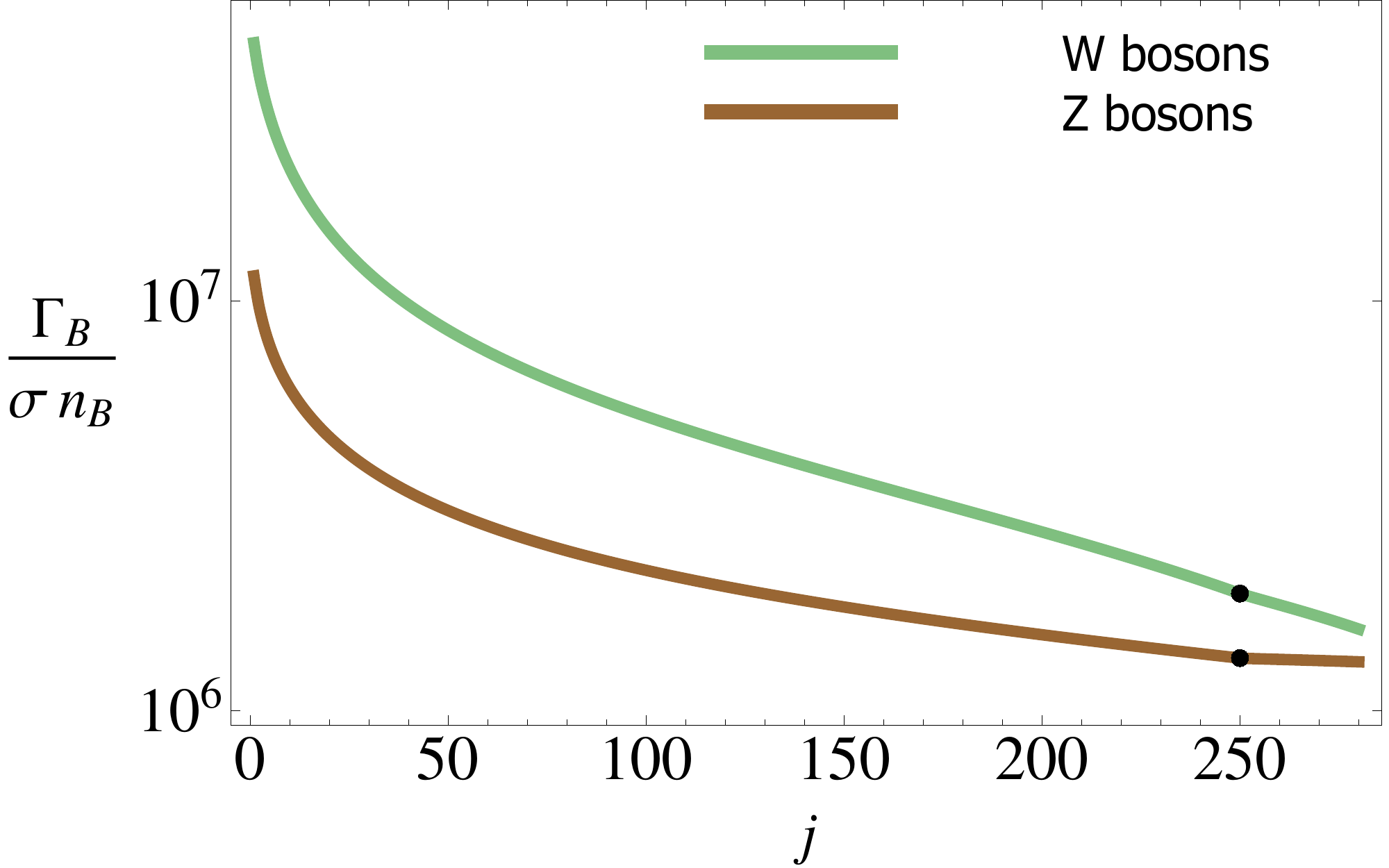}
\caption{(color online) Comparison between the decay channels $B\rightarrow F\bar F$ and the scattering channel $WW\rightarrow F\bar F$. }\label{gammafig}
\end{figure}

As shown in Fig.~\ref{brfig}, $\Delta M^B_{BR}$ and  $\Delta M^F_{BR}$ turn out to be completely negligible at all times before the onset of radiation domination, which justifies the use of Eq.~\eqref{nk}. 
\item
We verify \textit{a posteriori} the assumption $\Gamma_B \gg\sigma n_B $. The result showed in Fig.~\ref{gammafig} justifies the use of Eq.~\eqref{nk3}. 
\end{enumerate}

\section{Non-perturbative Higgs production in noncritical case} \label{appendix3}

In this appendix we derive an upper bound for the nonperturbative production of Higgs particles in noncritical Higgs inflation. In Fourier space, the Einstein-frame equation of motion for these perturbations reads
\begin{equation}\label{pertphi0}
\delta \chi''_k(t)+\left(K^2+\frac{V_{,\chi\chi}}{M^2}\right)\delta \chi_k(t)=0\,,
\end{equation}
with $K\equiv \frac{k}{aM}$ a rescaled momentum and the primes denoting derivatives with respect to a rescaled time $\tau=M t$.

A simple inspection of the potential in Fig.~\ref{potfig} suggests that the leading contribution to Higgs production should be associated with the region  $\vert \chi\vert \leq \chi_T$ in which the curvature of the potential ($V_{,\chi\chi}$) becomes negative \cite{Felder:2000hj}. Let us assume for simplicity that this curvature is constant. Denoting it by $-M_T^2$, we can rewrite Eq.~\eqref{pertphi0} as  
\begin{equation}\label{phiTeq}
\delta \chi''_k(t)-\Omega_T^2 \delta \chi_k(t)=0\,,\hspace{10mm} \Omega_T^2\equiv q_T^2-K^2\,,
\end{equation}
with $ q_T^2\equiv M_T^2/M^2\approx 0.067$ the numerical value of $M_T^2$ in units of the curvature of the potential at large field values.
The amplification of modes with $K<q_T$ is given by 
\begin{equation}\label{phiT}
\Delta \delta \chi_k^j\sim \exp\left(\Omega_T M \Delta t_j\right)\,,
\end{equation}
with $\Delta t_j$ denoting the time expended by the background Higgs field in the tachyonic region  $\vert \chi\vert \leq \chi_T$ for a given semioscillation $j$. Estimating this time via Eq.~\eqref{phievol}
\begin{equation}
M\Delta t \approx \frac{2\pi\chi_T}{\chi_e}\frac{1}{j}\,,
\end{equation}
we can rewrite Eq.~\eqref{phiT} as
\begin{equation}\label{deltaphi}
\Delta \delta \chi_k^j\sim e^{ A(j) j}\,, \hspace{10mm} A(j)\equiv\frac{2\pi\chi_T \Omega_T(j)}{\chi_e}
\end{equation}
The total amplification after a given number of semioscillations will be the result of the interference of the individual amplifications \eqref{deltaphi}. Since we are just seeking an upper bound on Higgs production, we will assume that the interference is fully constructive and that all the modes within the band $K< q_T$ grow at the maximum possible rate (i.e at the rate of the zero mode, $k=0$). 
\begin{equation}
A\approx \frac{2\pi\chi_Tq_T}{\chi_e}
\end{equation}
With these two assumptions, the occupation number of a given mode $k$ after $j$ semioscillations becomes
\begin{eqnarray}
n_k^j&\sim& \prod\limits_{j}^{} (\Delta \delta \chi_k^j)^2\simeq\exp\left(2 A \sum_j  j\right)\simeq e^{Aj^2}
\end{eqnarray}
where in the last step we have assumed a large number of semioscillations and approximated $\sum_j j=1/2 j(j+1)\simeq j^2/2$.

The total number of Higgs particles at their associated energy density after $j$ semioscillations is obtained by integrating over all the amplified modes $K< q_T$ (i.e, over $k< a M_T$). Taking into account that $a=j^{2/3}$, we obtain
\begin{eqnarray}
&&\hspace{-5mm} n_{\delta\chi}(j)=\frac{1}{2\pi^2 a^3}\int_0^{a M_T} d k\, k^2\, n_k 
\simeq \frac{q_T^3 M^3 }{6\pi^2} e^{Aj^2}\\
&&\hspace{-5mm}\rho_{\delta\chi}(j)=
\frac{1}{2\pi^2 a^3}\int_0^{a  M_T} d k\, k^2\,\vert \Omega_k\vert n_k 
\simeq \frac{q_T^4 M^4}{6\pi^2} e^{Aj^2} \label{rhoHpart}
\end{eqnarray}
Evaluating \eqref{rhoHpart} at the time in which the energy density of fermions starts dominating the expansion of the Universe ($j^*=250$), we get 
\begin{equation}
\rho_{\delta\chi}(j^*)\simeq 10^{-4} \rho_F(j^*),
\end{equation}
The contribution of Higgs particle production in noncritical Higgs inflation is completely negligible even with the extreme approximations performed in this appendix (namely constructive interference and maximum growing for all the modes.).

\section{Nonperturbative particle production in the critical case}\label{appendix4}

A proper treatment of particle creation in critical Higgs inflation would require the joint analysis of combined preheating and the tachyonic Higgs particle production at the bottom of the potential. Here we will simply try to derive some estimates on the separated processes. 
We start by applying the combined preheating formalism to the first few semioscillations ($j<10$) in which the quadratic approximation \eqref{potQ} still holds. The resulting evolution of the different energy densities\footnote{To avoid misunderstandings, we emphasize that the unit $M$ is computed with the values of $\xi$ and $\lambda$ associated to the critical case.} is shown in Fig.~\ref{rhocritfig}.  As in the noncritical case, the creation of particles is not efficient enough to dominate the expansion of Universe in such a short period of time.  Although the features of the potential could change the pattern of particle creation at later times, the energy density into fermions will never exceed\footnote{The quantity $\rho_\chi$ traces the maximum (total) energy of the system (it takes into account the expansion of the Universe but not the decay due to particle creation).} the energy of the Higgs field at $j\simeq 10$.
\begin{figure}
\centering
\includegraphics[scale=0.4]{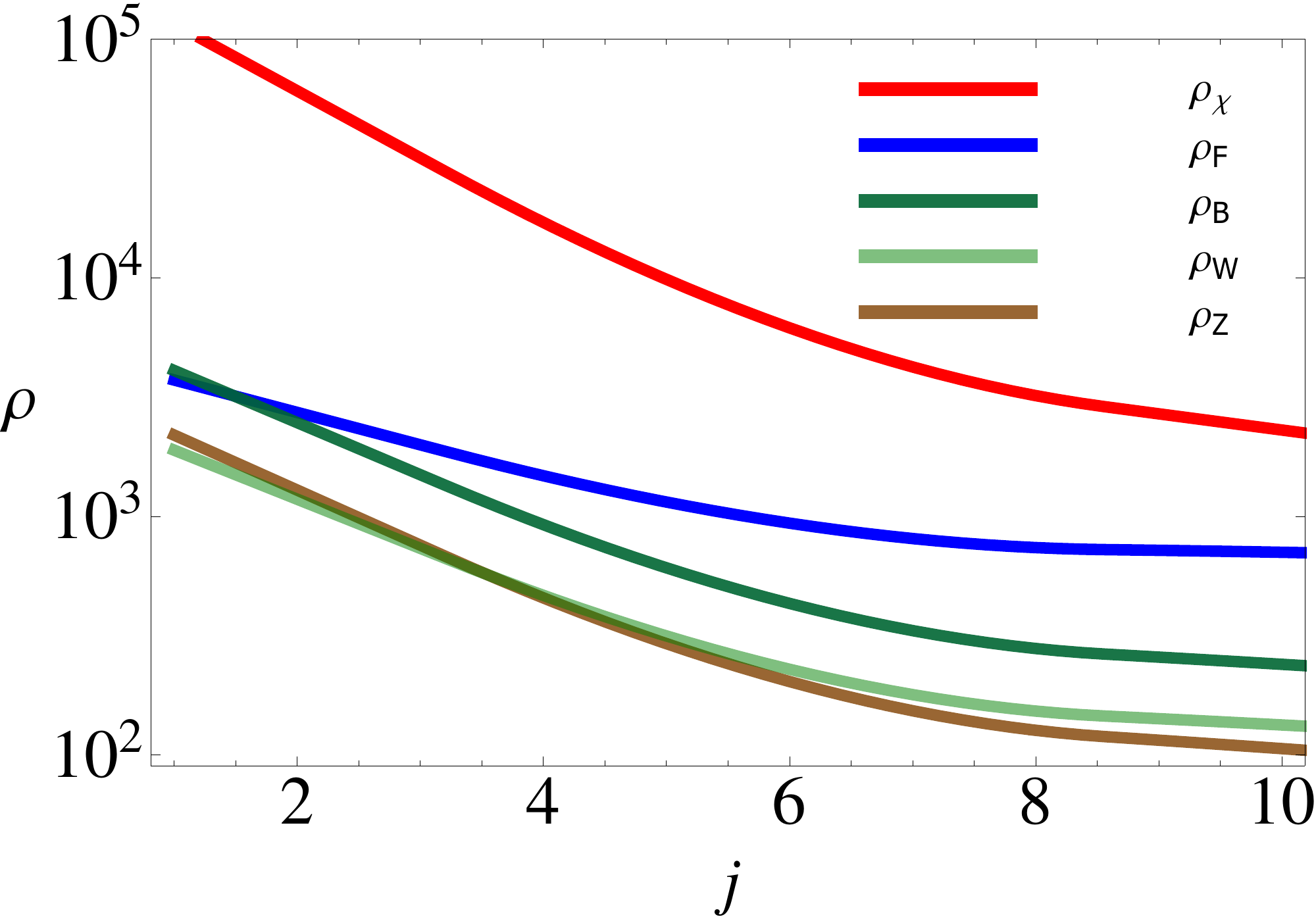}
\caption{(color online) Evolution of the different energy densities (in $M^4$ units) for the critical case ($\lambda=3\times 10^{-4}$, $\xi=15$, $g_1=0.44$, $g_2=0.53$).}\label{rhocritfig}
\end{figure}
\begin{figure}
\centering
\includegraphics[scale=0.4]{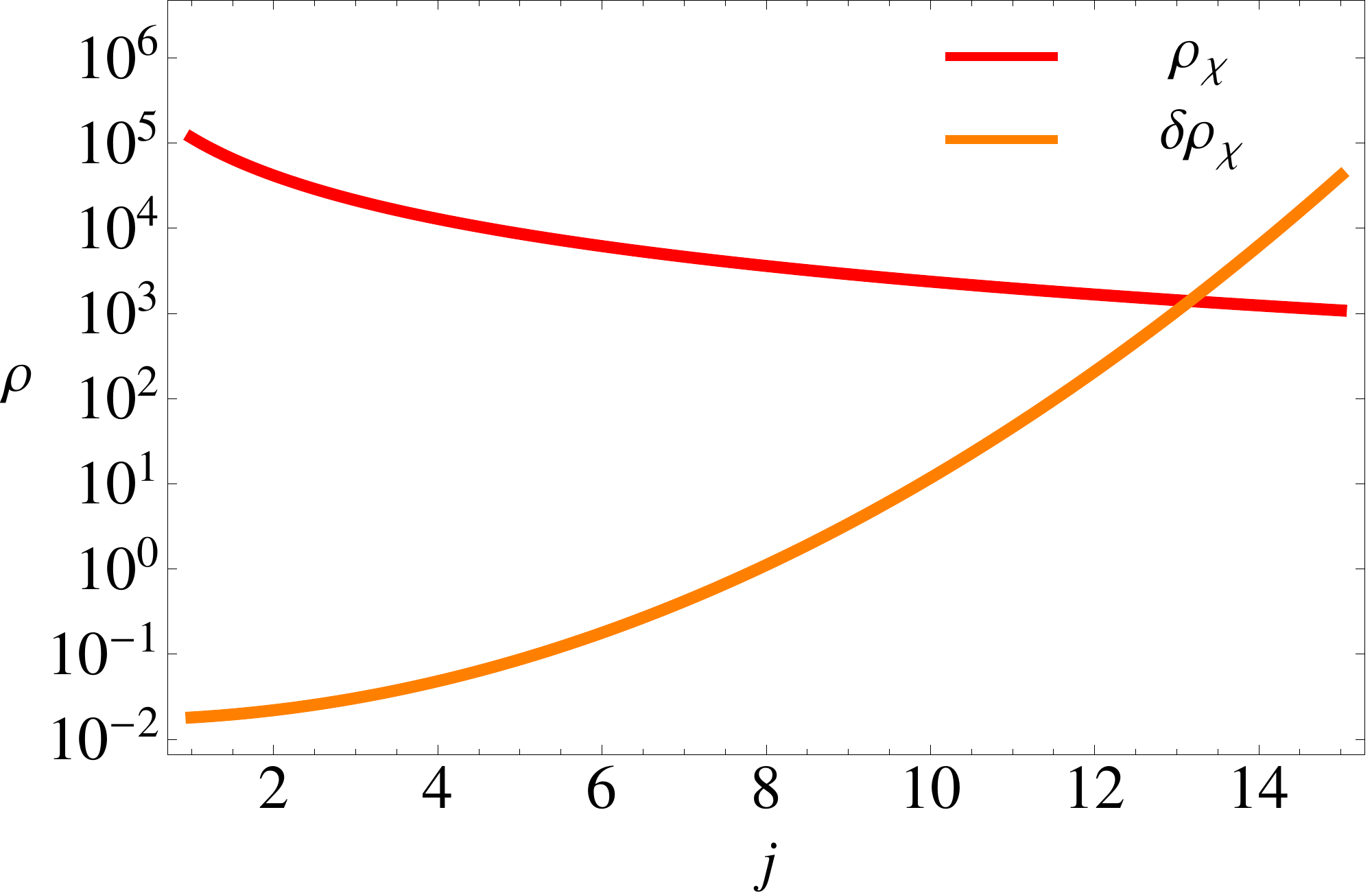}
\caption{(color online) Comparison between the energy density of the background and that into Higgs particles (both in $M^4$ units) as a function of the number of semioscillations $j$.}\label{rhotach}
\end{figure}

The production of Higgs particles by tachyonic instability can be estimated along the lines presented in Appendix \ref{appendix3}. In order to obtain a conservative bound for the maximum reheating temperature, we will assume that the interference between the different scattering is maximally constructive and that all the tachyonic modes grow at the maximum rate (i.e that the zero mode $k=0$). As shown in Fig.~\ref{rhotach}, the energy density into Higgs particles equals the energy into the background field in just $j\simeq 13$ semioscillations. The tachyonic production becomes the leading mechanism for particle creation in the noncritical case\footnote{Contrary to the noncritical case, the region in which the effective mass of the Higgs (i.e.~$V_{\chi\chi}$) becomes negative is quite large  [$\Delta\chi \sim {\cal O}(10^{-2}) M_P$)].}. Transforming this energy into the instantaneous radiation temperature of a plasma containing $g_*=68.25$ degrees of freedom, we obtain an upper bound for the reheating temperature in the critical case, namely
\begin{equation}
T^C_\textit{RH}<5 \times 10^{15}\,\text{GeV}\,.
\end{equation}
Since this temperature is smaller than the restoration temperature $T_{+}$, the shape of the potential remains unchanged and the system eventually finishes in the wrong minimum at large field values.

\end{appendix}

\end{document}